\documentclass[aps,pre,floatfix,twocolumn,nofootinbib,showpacs]{revtex4}
\usepackage{dcolumn}
\usepackage{bm}
\usepackage[dvips]{epsfig}
\usepackage{graphicx} \usepackage{amsmath} \usepackage{amssymb}
\newcommand{\comment}[1]{}
\newcommand{\BEQ}{\begin{equation}}
\newcommand{\EEQ}{\end{equation}}
\newcommand{\BEA}{\begin{eqnarray}}
\newcommand{\EEA}{\end{eqnarray}}

\renewcommand{\d}{{\rm d}}

\newcommand{\ec}{c_{\rm e}}

\renewcommand{\v}{\vec{v}}
\newcommand{\vnabla}{\vec{\nabla}}

\begin{document}

\title{Modeling gasodynamic vortex cooling \footnote{Published as 
A. E. Allahverdyan and S. Fauve, Phys. Rev. Fluids {\bf 2}, 084102 (2017). }}

\author{A. E. Allahverdyan$^{1)}$ and S. Fauve$^{2)}$}

\address{$^{1)}$ Yerevan Physics Institute, 2 Alikhanian Brothers
  street, 375036 Yerevan, Armenia, \\
  $^{2)}$Laboratoire de Physique Statistique, \'{E}cole Normale
  Sup{\'e}rieure, PSL Research University; UPMC Univ Paris 06,
  Sorbonne Universit\'{e}s; Universit\'{e} Paris Diderot, Sorbonne
  Paris-Cit\'{e}; CNRS; 24 Rue Lhomond, 75005 Paris, France}

\begin{abstract} 
  We aim at studying gasodynamic vortex cooling in an analytically
  solvable, thermodynamically consistent model that can explain
  limitations on the cooling efficiency. To this end, we study a
  angular plus radial flow between two (co-axial) rotating permeable
  cylinders. Full account is taken of compressibility, viscosity and
  heat conductivity.  For a weak inward radial flow the model
  qualitatively describes the vortex cooling effect|both in terms of
  temperature and of decrease of the stagnation enthalpy|seen in short
  uniflow vortex (Ranque) tubes. The cooling does not result from
  external work, and its efficiency is defined as the ratio of the
  lowest temperature reached adiabatically (for the given pressure
  gradient) to the actually reached lowest temperature. We show that
  for the vortex cooling the efficiency is strictly smaller than 1,
  but in another configuration with an outward radial flow, we found
  that the efficiency can be larger than 1. This is related to both
  the geometry and the finite heat conductivity.

\end{abstract}


\comment{Vortex dynamics (fluid flow), 47.32.C- Compressible flows,
  47.40.-x Refrigeration, 07.20.Mc Irreversible thermodynamics,
  05.70.Ln thermodynamics, 05.70.-a }

\maketitle

\section{Introduction}
\label{intro}

Air swirling through a cylindrical tube achieves temperature
separation: next to the swirling axis the temperature is lower than
the input temperature $T_0$, while far from the axis it is higher than
$T_0$. This is the vortex cooling-heating effect discovered by
G. Ranque more than 80 years ago \cite{ranque,hilsch}; see
\cite{graham,gutsol,xue,guliguli} for reviews. A temperature
separation without cooling was observed also in highly-pressurized
water \cite{balmer}. An overall cooling (both output temperatures lower
than $T_0$) was seen for certain vortex tubes \cite{finko}.

Coolers based on the Ranque effect are convenient in specific
applications, e.g. because they do not have moving parts. However,
their efficiency is smaller than 1. Much effort was devoted to
increase it, but the best efficiency is still $\simeq 0.6$
\cite{gutsol,guliguli}.

The flow inside the Ranque tube is highly complex: it is essentially
three-dimensional and turbulent. Two configurations have been used:
counter-flow and uniflow \cite{graham,gutsol,xue,guliguli}. In
counter-flow vortex tubes the output flows are collected from two
different ends of the cylinder, the cold air is extracted from a small
aperture around the axis and close to the injection point, while the
hot air comes out from the opposite end of the cylinder without being
collimated close to the axis \cite{graham,gutsol,xue,guliguli}.  Such
tubes have both radial \cite{graham,gutsol,farouk} and axial
temperature separation \cite{eckert,dinesh,seca,xue,dubno}. In the
uniflow situation the air is injected circumferentially at one end of
the tube, and both output flows are collected from the opposite end
\cite{graham,tai_uniflow}. This radial temperature separation takes
place close to the injection point of the air \cite{leites}.

The full theory of the Ranque effect is elusive; there are several
different approaches that attempt to describe the complex
three-dimensional flow inside of the tube
\cite{adiabat_aleks,adiabat_canada,adiabat_hashem,crit_adiabat,adiabat_dutch,
  adiabat_kalashnik,dorn,penge,knoer,deemter,reynolds}. In particular,
it is unclear what are the minimal ingredients needed to describe the
effect.

Given the complexity of the original Ranque effect, and the necessity
of understanding general limitations on the efficiency of gasodynamic
cooling, it is desirable to come up with a simpler cooling set-up,
where the complexities of the original Ranque effect are deliberately
omitted. The quest for such a simplification was already considered in
Ref.~\cite{savino}, where Savino and Ragsdal reported on an
experimental realization of a short uniflow tube, where the flow is
injected from the surface via permeable rotating wall, and the colder
air is collected from the axis \cite{savino}. Axial separation of
temperature is absent, and the whole outgoing flow is cooled
\cite{savino}.

\begin{figure*}
\begin{center}
{\Large\bf Figures}
\end{center}
\includegraphics[width=10.4cm]{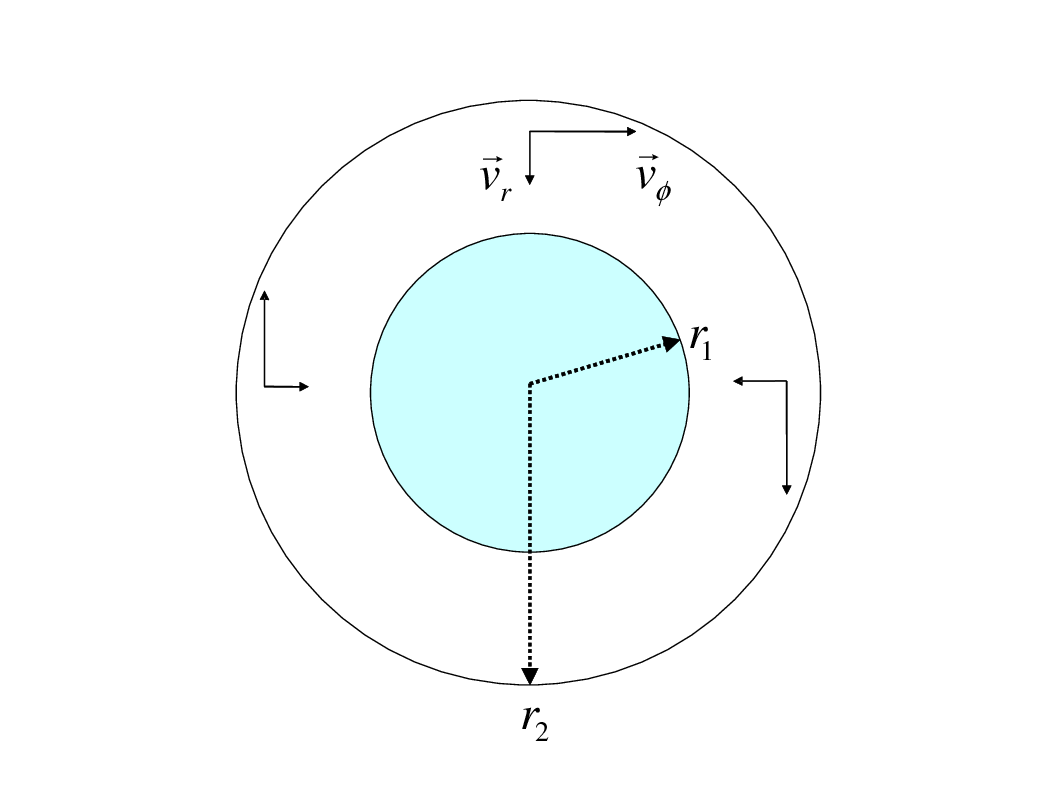}
\caption{Cross-section of the flow: two coaxial permeable
  cylinders with radii $r_1$ and $r_2$ rotate with prescribed
  speeds. Solid vectors $\vec{v}_\phi={v}_\phi\vec{e}_\phi$ and
  $\vec{v}_r={v}_r\vec{e}_r$ refer to the velocity components of the
  flow for $r_1<r<r_2$. Here the radial flow is directed from the
  outer (larger) cylinder to the inner (smaller) cylinder. Now
  $|\vec{v}_r|<|\vec{v}_\phi|$, since the radial flow has to be smaller
  for cooling.  }
\label{f0}
\end{figure*}

Guided by this experiment, we aim at understanding the phenomenon of
vortex|and more general gasodynamic|cooling on a possibly simple
theoretical model. We focus on a compressible angular flow between two
rotating cylinders plus a radial motion via permeable cylinder walls;
see Fig.~\ref{f0}. We work with a compressible flow, because
experimental angular velocities are nearly sonic
\cite{gutsol,savino}. Moreover, once we are interested by
thermodynamic aspects (i.e. cooling efficiency), it is desirable to
work in the compressible situation, where the thermodynamic
description of the flow is complete and consistent\footnote{The
  incompressible limit is singular from the viewpoint of
  thermodynamics \cite{ottinger}. Despite of the widespread usage of
  this limit, its consistent thermodynamics was developed only
  recently \cite{ottinger}.}. We assume that the flow is viscous,
because (according to the Bernoulli's theorem) the adiabatic motion of
the fluid does not predict cooling in terms of stagnation
enthalpy. However, precisely such cooling is observed experimentally
\cite{savino} \footnote{Thus adiabatic theories of the Ranque effect
  \cite{adiabat_aleks,adiabat_canada,adiabat_hashem,adiabat_dutch,
    adiabat_kalashnik} do not describe the full cooling effect
  \cite{crit_adiabat}.}. Hence viscosity is important \cite{guliguli},
and then heat-conductivity is to be accounted for simultaneously with
viscosity, because the Prandtl number of air is close to one both in
laminar and turbulent regimes.

Our first result is that in the stationary regime of a weak radial
flow and a quasi-solid angular (vortical) motion, the model predicts
cooling both in terms of thermodynamic temperature and stagnation
enthalpy. The efficiency of this cooling is smaller than 1.  This
qualitatively agrees with experiments \cite{savino}. The agreement is
achieved by using effective (turbulent or eddy) values of viscosity
and heat-conductivity in the laminar flow model. Such an approach is
well-known \cite{landau}. Using turbulent diffusivites is crude but
provides qualitative results \cite{knoer,deemter,reynolds} that allow
a theoretical understanding of the cooling effect.

The model predicts a stronger cooling effect for (radially) outward
flow of fluid. The unique feature of this effect is that its cooling
efficiency is larger than 1, i.e. the pressure gradient is employed
more efficiently than for the adiabatic process. This effect agrees
with the second law, and it is possible due to heat-conductivity.

Scenarios of cooling studied here do not amount to refrigeration,
i.e. they are not achieved by investing an external work. Naturally,
they do not result either from boundaries maintained at low
temperature by a heat bath; to ensure this we need to pay a special
attention to boundary conditions. Hence cooling efficiency is defined
as the ratio of the lowest temperature reached adiabatically (for the
given pressure gradient) to the actually reached lowest temperature.

Cylindrical vortices with radial flow were already studied in
Refs.~\cite{dorn,penge,deemter,perl,rott,bell,kolesov,
  tilton,terrill,terrill_1,polonius}, but the problem of finding
cooling scenarios with proper boundary conditions was (to our
knowledge) not posed.  Dornbrand \cite{dorn} and later on Pengelley
\cite{penge} studied the problem precisely having the same purpose as
we: to get a solvable model for vortex cooling. But they did not
account for boundary conditions and heat conduction and thus did not
obtain proper cooling. Pengelley proposed a necessary condition for
cooling that relates to the work done by viscous forces
\cite{penge}. Below we show that under certain additional limitations
this condition is indeed able to produce
cooling. Refs.~\cite{knoer,deemter,reynolds} used simplified
turbulence theories of various types that account for radial heat
conductivity and viscous vortex motion. A related, but more complete
turbulent theory that also accounts for axial motion was given in
Ref.~\cite{perl}. Refs.~\cite{rott,terrill,terrill_1,kolesov} focus on
a laminar flow in the incompressible limit (but they account for axial
motion).  As our analysis shows, compressibility does not need to be
large, but retaining it|and hence allowing for the proper coupling
between thermodynamics and mechanics|is necessary for the proper
theoretical description of cooling.  Ref.~\cite{bell} did not employ
the incompressible limit, but studied the problem without the outer
cylinder. Several studies on convective heat transfer between
concentric cylinders are reviewed in \cite{kumar,arab,kameroon}.

The paper is organized as follows. Next section defines the problem
and sets notations and dimensionless parameters. There we also discuss
general limitations (in particular, on the cooling efficiency) imposed
by the first and second laws. Section \ref{definition} focuses on the
definition of cooling, which is not trivial (especially for permeable
walls) and thus demands clarifications. Cooling scenarios of inward
radial flow are studied in section \ref{inward}. The extent to which
this scenario agrees with experiments is discussed in section
\ref{relo}.  Section \ref{semi} discusses the cooling of outward
radial flow and shows that its efficiency is larger than one.
Concluding remarks are given in the last section. Several technical
questions are relegated to Appendices.

\comment{
The second aspect is important, since the second law
does limit the cooling efficiency, and without understanding of
these limits in a sufficiently simple model, one can run into
contradictory statements that the Ranque effect does not agree with
the second law. }

\section{The model}
\label{notations}

\subsection{Navier-Stokes equation}

The flow between two rotating concentric cylinders is described via
cylindric coordinates $(r,\phi,z)$, and $\vec{v}=(v_r,v_\phi,v_z)$ are
components of the velocity. We assume that all the involved quantities
depend only on $r$, e.g. $\vec{v}=\vec{v}(r)$. We also assume that
$v_z=0$, since in the context of our problem it is useless to keep
$v_z\not =0$, if it is a function of $r$ only. A schematic
representation of the flow is displayed in Fig.~\ref{f0}.

In the stationary regime the Navier-Stokes equations for $v_r$ and
$v_\phi$ read \cite{landau}:
\begin{align}
  \label{eq:1}
&  \rho (v_r\frac{\d v_r}{\d r} -\frac{v_\phi^2}{r})=-\frac{\d p}{\d r}
+(\zeta+\frac{4\eta}{3})\,\frac{\d}{\d r}\left[\,\frac{1}{r}\, \frac{\d
  (rv_r)}{\d r}\,\right] ,\\
\label{eq:3}
&\rho(v_r\frac{\d v_\phi}{\d r}+\frac{v_rv_\phi}{r})=
\eta\left(\, \frac{1}{r} \frac{\d}{\d r}\left[r\frac{\d v_\phi}{\d r} 
\right]-\frac{v_\phi}{r^2}\,\right),
\end{align}
where $p$ is pressure, $\eta$ and $\zeta$ are viscosities, $\rho$ is
the mass density; see Table~\ref{tab1}. We assume that $\eta$ and
$\zeta$ are constants, i.e. they do not depend on $p$, $\rho$ or
$T$. Conservation of mass reads ($\vnabla$ is the gradient):
\begin{eqnarray}
  \label{eq:30}
&&\partial_t\rho+\vnabla (\rho \v)=r^{-1}\frac{\d}{\d r}(\rho rv_r)=0, \\  
  \label{eq:4}
&&  \rho r v_r=c={\rm const},
\end{eqnarray}
where $c$ (a positive or negative constant) characterizes the radial
flow. Eqs.~(\ref{eq:3}, \ref{eq:4}) transform to
\begin{eqnarray}
  \label{eq:5}
  \eta r^2\,\frac{\d^2 v_\phi}{\d r^2}+  (\eta-c) r\, \frac{\d v_\phi}{\d r}
-(\eta+c)v_\phi=0.
\end{eqnarray}
This is a homogeneous equation linear in $v_\phi$. Its two independent
solutions are obtained by putting $v_\phi\propto r^{a}$ into
(\ref{eq:5}). The latter produces a quadratic equation for $a$. This
equation has two solutions $a=-1$ and $a=1+\frac{c}{\eta}$.

We now impose boundary conditions on (resp.) inner and outer cylinder
\begin{eqnarray}
  \label{eq:6.2}
v_1\equiv v_\phi(r_1), \qquad v_2\equiv v_\phi(r_2),
\end{eqnarray}
where $r_2>r_1$. Then (\ref{eq:5}) and (\ref{eq:6.2}) are solved as a
linear combination of $a=-1$ and $a=1+\frac{c}{\eta}$
solutions of (\ref{eq:5}):
\begin{eqnarray}
  \label{eq:6}
&&  v_\phi(r)=v_2\,\hat{v}_\phi(x), \qquad x\equiv r/r_2, \\
\label{eq:6.0}
&& \hat{v}_\phi(x)=
[(1-\alpha)x^{-1} + \alpha x^{1+\kappa}  ],\\
  \label{eq:6.1}
&& \kappa \equiv \frac{c}{\eta}, \qquad
\alpha\equiv \frac{1-(v_1r_1)/(v_2r_2)}{1-(r_1/r_2)^{2+\kappa}},
\end{eqnarray}
where we introduced the dimensionless coordinate $x$; see
Table~\ref{tab1}. Eq.~(\ref{eq:6.0}) is a weighted sum of two
contributions: a potential vortex $1/x$ and a quasi-solid vortex
$x^{1+k}$. The weight $\alpha$ can hold both $\alpha>1$ and
$\alpha<0$.

\begin{table*}[ht]
  \caption{ Variables and parameters. Dimensionless
    quantities are indicated by $^*$. }
\begin{tabular}{|c|c|c|}
  \hline
Variable/Parameter   & Defined in Eq. & Description    \\
  \hline  \hline
  $r_2>r_1$   & (\ref{eq:6.2}) & Radii of the coaxial cylinders   \\
  \hline
  $x=r/r_2$~~$^*$   & (\ref{eq:6}) & Dimensionless radial distance   \\
  \hline
  $x_0=r_1/r_2$~~$^*$   & (\ref{taron}) & Dimensionless ratio of the radii   \\
  \hline
  $v_\phi$ & (\ref{eq:1},\ref{eq:3}) & Angular velocity \\
  \hline
  $v_2=v_\phi(r_2)$, $v_1=v_\phi(r_1)$   & (\ref{eq:6.2}) 
                                    & Angular velocities of the
                                      coaxial cylinders    \\
  \hline
  $v_r$ & (\ref{eq:1},\ref{eq:3}) & Radial velocity \\
  \hline
  $w=xv_r/v_2$~~$^*$ & (\ref{eq:34}) & Dimensionless radial velocity \\
  \hline
  $\rho$  & (\ref{eq:1},\ref{eq:3})  & Mass density. Under normal
  conditions for air: $\rho=1.2$ kg/${\rm m}^3$ \\
  \hline
  $p$  & (\ref{eq:1},\ref{eq:3})        & Pressure \\
  \hline
  $\rho\epsilon$        & (\ref{hu})  & Internal energy density \\
  \hline
  $\rho s$    & (\ref{eq:26_1}, \ref{eq:11}) & Internal entropy density \\
  \hline
  $\lambda$ & (\ref{hu}) & Heat-conductivity. Molecular value for air: \\
            &            & $\lambda_{\rm mol}=0.02$ $\frac{\rm J}{\rm m\,s\,K}$.  
            For
            turbulent value see section \ref{compo}. \\
  \hline  
  $T$                   & (\ref{hu}) & Temperature measured in Kelvins \\
  \hline
  $\widehat{T}=\lambda T/(v_2^2 \eta)$~~$^*$ 
                      & (\ref{eq:34}) & Dimensionless temperature \\
  \hline
  $c=\rho v_rr$   & (\ref{eq:4}) & Radial flow. $c$ is a constant with
                                   this model \\
  \hline
  $c_p$   & (\ref{eq:4}) & Isobaric heat capacity. For air: $c_p=10^3$
                                     $\frac{\rm J}{\rm kg\, K}$ \\
  \hline
  $\hat{c}_p$~~$^*$   & (\ref{ole}) &  Dimensionless isobaric
                              heat-capacity; $\hat{c}_p/(\hat{c}_p-1)=c_p/c_v$ \\
  & & is the ratio of isobaric and isochoric heat-capacities\\
  \hline
  $\eta$, $\zeta$   & (\ref{eq:1},\ref{eq:3}) & Viscosities. 
                        Molecular value of air: \\
  &               &     $\eta_{\rm mol}=1.8 \times 10^{-5}$
              $\frac{{\rm kg}}{{\rm m\,\, s}}$. For turbulent value see section 
              \ref{effo}.  \\
  \hline
  $\chi=\eta/\zeta$~~$^*$   & (\ref{eq:34.1}) & Ratio of viscosities
  \\
\hline
$\alpha=\frac{1-(v_1/v_2)\,x_0}{1-x_0^{2+\kappa}}$ ~~$^*$ & (\ref{eq:6.1}) & The
weight of the quasi-solid vortex in the angular motion \\
  \hline 
  $\kappa=c/\eta$~~$^*$  & (\ref{eq:34.0}) &  $|\kappa|$ is the
  Reynolds number related to the radial flow   \\
  \hline  
    $b=cc_p/\lambda$~~$^*$   & (\ref{eq:34.0}) & $|b|$ is the Peclet number \\
  \hline
  $\beta$~~$^*$   & (\ref{eq:34.1},\ref{bora}) &  
  $|\beta|$ is the analogue of the Reynolds number \\
  & &  related to the radial flow of energy  \\
  \hline
${\rm Pr}=b/\kappa= \eta c_p/\lambda $~~$^*$ & (\ref{eq:41})& Prandtl number\\
  \hline
$\hat{U}$~~$^*$ & (\ref{stop})& Stagnation enthalpy \\
\hline
\end{tabular}
\label{tab1}
\end{table*}

\subsection{Energy equation}

The fluid energy equation reads \cite{landau}
\begin{gather}
  \label{hu}
  \partial_t (\frac{\rho \,\v^2}{2} +\rho\varepsilon   )
+\vnabla
  [\,\rho\v (\frac{\v^2}{2} +\varepsilon  ) +p\,\v
+\vec{\mu}-\lambda\vnabla T  \, ]=0,
\end{gather}
where $\frac{\rho \v^2}{2} +\rho\varepsilon$ is the energy density
(kinetic energy plus internal energy),
$\rho\v (\frac{\v^2}{2} +\varepsilon )$ is the advective energy flux,
$p\v$ is the pressure-driven energy flux, $T$ is absolute temperature
(measured in Kelvins), $\vnabla (\lambda \nabla T)$ is the heat flux
with heat conductivity $\lambda$ (we assume that $\lambda$ does not
depend on $p$, $\rho$ and $T$), $\mu_k=-\sum_jv_j\sigma_{jk}$ is the
energy flow due to viscosity, and $\sigma_{jk}$ is the stress tensor
\cite{landau}.

With the assumptions and in the stationary regime the energy flux is
$\ec/r$, where $\ec$ is a constant [cf.~(\ref{eq:4})]:
\begin{eqnarray}
\label{bora}
&& \ec=c E
-rv_r\sigma_{rr}-rv_\phi\sigma_{r\phi}
-\lambda r\frac{\d T}{\d r},\\
  \label{eq:9}
&& E=\frac{v_r^2+v_\phi^2}{2}+ \varepsilon+\frac{p}{\rho}, \\
  \label{eq:10}
&&  \sigma_{rr}=2\eta \frac{\d v_r}{\d r}+(\zeta-\frac{2\eta}{3})\,
\frac{1}{r}\,\frac{\d (rv_r)}{\d r},\\
&&  \sigma_{r\phi}=\eta ( \frac{\d v_\phi }{ \d r}-\frac{ v_\phi }{r}) .
  \label{eq:100}
\end{eqnarray}
Here $E$ is the full energy (kinetic + internal + potential) per unit
of mass; $-\lambda\, \frac{\d T}{\d r}$ is the heat flux due to the
radial temperature gradient; $\sigma_{rr}$ and $\sigma_{r\phi}$ are
the components of the stress tensor \cite{landau}, $v_r\sigma_{rr}$
($v_\phi\sigma_{r\phi}$) is the rate of radial (angular) work done by
viscous forces. Eqs.~(\ref{bora}--\ref{eq:100}) express the first law
for the radial flow.

Eqs.~(\ref{eq:1}) and (\ref{bora}) become closed after specifying the
thermodynamic state equation; we choose it by assuming that the fluid
holds the ideal gas laws [see Appendix \ref{idol}]:
\begin{eqnarray}
  \label{ideal}
&&p={R}\rho T/{\mu} , \\
  \label{rashid}  
  \label{eq:12}
&& (\rho\varepsilon+{p})/{\rho} =c_pT, \\
\label{ole}
&& c_p=\hat{c}_p\, R/\mu,
\end{eqnarray}
where $c_p>0$ is the (constant) heat capacity at fixed pressure, and
$\hat{c}_p$ is a dimensionless number of order 1 (e.g. $\hat{c}_p
\approx 3.5$ for air); see Table~\ref{tab1}. $R=8.314$ J/K is the gas
constant and $\mu$ is the molar mass (29 g for air). 


\subsection{Dimensionless parameters and variables}

Employing (\ref{eq:10}--\ref{ole}) in (\ref{bora}), and (\ref{ideal},
\ref{eq:12}) in (\ref{eq:1}), we end up with the following
dimensionless form of (respectively) (\ref{bora}) and (\ref{eq:1}):
\begin{eqnarray}
&&  (\frac{\kappa}{2}+1) \hat{v}_\phi^2 - x
  \hat{v}_\phi\hat{v}'_\phi
  +b\hat{T}-x\hat{T}' -\beta ~~~~~~~ \nonumber\\
  \label{s2}
&& +
  (\frac{\kappa}{2}+2)\frac{w^2}{x^2}  -(\chi+\frac{4}{3}) \frac{ww'}{x}
  =0,\\
&&  (\chi+\frac{4}{3})w''-(\kappa+\chi+\frac{4}{3})\frac{w'}{x} 
  +\frac{\kappa w}{x^2}
\nonumber  \\
  \label{s4}
&&  +\frac{\kappa\hat{v}_\phi^2 }{w}-\frac{b\, x}{\hat{c}_p}\,(\hat{T}/{w})'
=0, 
\end{eqnarray}
where $x=\frac{r}{r_2}$ [cf. (\ref{eq:6}, \ref{eq:6.1})], 
prime means $\frac{\d}{\d x}$, e.g.
\begin{eqnarray}
  \label{eq:61}
  \hat{v}_\phi'=\frac{\d \hat{v}_\phi}{\d x},
\end{eqnarray}
and where we introduced [cf.~Table~\ref{tab1}]:
\begin{eqnarray}
  \label{eq:34}
&& \hat{T}=\frac{\lambda }{v_2^2\eta}T, ~~~~ w=\frac{xv_r}{v_2} \\
  \label{eq:34.0}
&& b=\frac{cc_p}{\lambda}, ~~~~ \kappa=\frac{c}{\eta}, \\
  \label{eq:34.1}
&& \chi=\frac{\zeta}{\eta}, ~~~~ \beta=\frac{c_{\rm e}}{\eta v_2^2}.
\end{eqnarray}
Here $|\kappa|$ is the Reynolds number related to the radial flow,
while $b/\kappa$ is the Prandtl number. These and other dimensional
and dimensionless parameters of the systems are discussed in
Table~\ref{tab1}.  The angular Mach number ${\rm Ma}$ of the outer
cylinder reads via the above parameters as
\begin{eqnarray}
  \label{mach}
{\rm Ma}=\sqrt{\frac{(\hat c_p-1)\kappa}{b\hat T}}=\frac{|v_2|}{v_{\rm
    sound}},  
\end{eqnarray}
where $v_{\rm sound}=\sqrt{\frac{c_pT}{(\hat c_p-1)}}=\sqrt{\frac{\hat
    c_p}{\hat c_p-1}\, \frac{p}{\rho}}$ is the speed of sound for the
ideal gas; see (\ref{ideal}) and \cite{landau}. The constant $\beta$
in (\ref{s2}) can be related to $ \hat{T}'(1)$ via $\hat{v}_\phi'(1)$
[see (\ref{eq:6.0})]:
\begin{gather}
 \hat{T}'(1)=
-\beta+b\hat{T}(1)+(2+\frac{\kappa}{2})w^2(1)
-(\chi+\frac{4}{3}) w(1)w'(1)\nonumber\\
-\alpha(2+\kappa)+2+\frac{\kappa}{2}.
\label{eq:15}
\end{gather}

\subsection{Scaling of temperature}

The scaling over $v_2$ employed in (\ref{eq:34}) and in (\ref{eq:6})
does have a physical meaning, since below we show that cooling
(i.e. temperature decrease) relates to the angular motion of the
cylinders. The choice of the dimensionless temperature $\hat T$ is
also a reasonable one because, using typical values of experiments
\cite{gutsol,savino}, we find $\hat{T}\sim 1$. Indeed, using
(\ref{eq:34}) we obtain
\begin{eqnarray}
  \label{eq:41}
  T(r_2)=\hat{T}(1){v_{\rm sound}^2}\,\,{\rm Pr}\,\, {\rm Ma}/{c_p},
\end{eqnarray}
where $v_{\rm sound}$ is the sound velocity, and where ${\rm Pr}\equiv
\eta c_p/\lambda$ and ${\rm Ma}=v_\phi(r_2)/v_{\rm sound}$ are the (resp.)
Prandtl and Mach numbers; cf.~(\ref{mach}). Recall that
$\hat{T}(1)=\hat{T}(x=1)$, where $x=r/r_2\leq 1$ is the dimensionless
length. For air we take in (\ref{eq:41}): $c_p=10^3$ J/(kg\, K) and
$v_{\rm sound}=3.31\times 10^2$ m/s. In vortex cooling experiments,
the input air has ${\rm Ma}\sim 1$ \cite{xue,gutsol}. Also ${\rm
  Pr}\sim 1$ holds for air \footnote{This relation holds both for the
  laminar regime and in the fully developed turbulence regime
  \cite{landau,shirokov}. For the former (latter) we employ molecular
  (turbulent) values of heat-conductivity and viscosity; see section
  \ref{relo}. }.  For an inward flow $c<0$, the input temperature is
$T(r_2)$. We get for it: 
\begin{eqnarray}
  \label{star}
T(r_2)\simeq\hat{T}(1)\times 100.  
\end{eqnarray}
Thus room temperature $T(r_2)\simeq 300$ K means $\hat{T}(1)\simeq 3$.

However, the above scaling of the dimensionless temperature $\hat{T}$
is not applicable for $v_2\to 0$. Then we should change $v_2\to v_1$
in (\ref{eq:34}, \ref{eq:34.1}) and take instead of (\ref{eq:6})
\begin{eqnarray}
{v}_\phi(r)=v_1 \hat{v}_\phi(x), \quad 
\hat{v}_\phi(x)=\frac{x^{-1}-x^{1+\kappa}}{x_0^{-1}-x_0^{1+\kappa}},
\quad
x_0=\frac{r_1}{r_2}.  
\label{taron}
\end{eqnarray}

\subsection{First law}

Using (\ref{eq:34}--\ref{eq:34.1}), the energy balance (\ref{bora})
can be written in terms of dimensionless, local rates of energy
$\widehat{E}$, radial work $\widehat{W}_r$, angular work
$\widehat{W}_\phi$ and heat $\widehat{Q}$:
\begin{gather}
\widehat{E}(x)=\frac{c}{\eta v_2^2}[\, 
\frac{v_r^2(r)+v_\phi^2(r)}{2}+c_p T(r)
\,  ]\nonumber\\
\label{energy}
= b \hat{T}+\frac{\kappa}{2}(\hat{v}_\phi^2+ \frac{w^2}{x^2}),\\
\label{work_tangential}
\widehat{W}_\phi(x)= -\frac{r}{\eta
  v_2^2}\,v_\phi(r)\sigma_{r\phi}(r)
=
\hat{v}_\phi(x)[\,\hat{v}_\phi(x)-x\hat{v}_\phi'(x) ], \\
\label{work_radial}
\widehat{W}_r(x)= -\frac{r}{\eta v_2^2}\, v_r(r)\sigma_{rr}(r)
=\frac{2w^2}{x^2} -(\chi+\frac{4}{3})\frac{ww'}{x},\\
\label{heat}
\widehat{Q}(x) = -\frac{\lambda r}{\eta v_2^2}T'(r_1)= -x\hat{T}'(x),
  \end{gather}
where we employed (\ref{eq:34}--\ref{eq:34.1}) and
(\ref{eq:6.2}--\ref{eq:6.1}).

The first law reads from (\ref{bora}):
\begin{eqnarray}
\label{1law}
  \Delta
  \widehat{E}+\Delta\widehat{W}_r+\Delta\widehat{W}_\phi+\Delta
  \widehat{Q}=0, 
\end{eqnarray}
where
$\Delta \widehat{E}\equiv\widehat{E}(1)-\widehat{E}(\frac{r_1}{r_2})$
{\it etc}. Note that $\Delta (\widehat{W}_r+\widehat{W}_\phi)>0$ means
that the system does work on the external sources which immerse the
fluid into the system and rotate the cylinders, i.e. the work is
extracted \footnote{Note from (\ref{hu}) that the integral
  $\int_{\cal \partial V} \d \vec{s}\,\vec{\mu}$, over a closed
  surface ${\cal \partial V}$ is the work done by viscosity forces on
  the substance enclosed into ${\cal \partial V}$. The sign of the
  work is determined as follows: with the normal vector $\vec{s}$ of
  ${\cal \partial V}$ pointing outside, the integral contributes into
  $ -\partial_t \int_{\cal V} \,{\rm d}V\, (\frac{\rho \v^2}{2}
  +\rho\varepsilon )$; see (\ref{hu}). Hence a positive
  $\int_{\cal \partial V} \d \vec{s}\,\vec{\mu}$ means that the fluid
  in ${\cal V}$ does work on external sources.}. We stress that this
model of cooling is not completely autonomous, since it contains
moving boundaries. The condition
$\Delta\widehat{W}_r+\Delta\widehat{W}_\phi\geq 0$ means that cooling
(if it is shown to exist) is not due to external forces that move
boundaries.

Let us also give the dimensionless form of the stagnation enthalpy: 
\begin{eqnarray}
\label{stop}
\hat U=\frac{1}{v_2^2}(\frac{ \v^2}{2} +\frac{p}{\rho}
+\varepsilon) =\frac{\hat{v}_\phi^2}{2}+\frac{w^2}{2x^2}+\frac{b\,
  \hat{T}}{\kappa},
\end{eqnarray}
which differs from (\ref{energy}) by the factor $c/\eta$ only. 

\subsection{Angular work}

The work (\ref{work_tangential}) done by rotating cylinders
can be calculated in a closed form from (\ref{eq:6}, \ref{eq:6.1},
\ref{eq:100}):
\begin{gather}
\Delta\widehat{W}_\phi=
2(1-\alpha)-\kappa\alpha
-\left[\frac{1-\alpha}{x_0}+\alpha x_0^{1+\kappa}\right]\times\nonumber\\
\left[\frac{2(1-\alpha)}{x_0}-\alpha\kappa x_0^{1+\kappa}\right], \quad 
x_0\equiv \frac{r_1}{r_2}.
  \label{vagr}
\end{gather}
Now radially outward flow means $\kappa \geq 0$ or $c\geq 0$; see
(\ref{eq:34.1}). For this case we checked numerically that
$\Delta{W}_\phi\leq 0$, i.e. the cylinders always invest work. In
particular, $\Delta\widehat{W}_\phi=2(1-\alpha)^2(1-x_0^{-2})<0$ for
$\kappa=0$.  For inward flow $\kappa < 0$ there are situations, where
$\Delta{W}_\phi> 0$, i.e. the work is extracted.  Note from
(\ref{eq:6}, \ref{eq:6.0}) that $\kappa<0$ means that the angular
velocity $v_\phi/r$ is a decreasing function of $r$.

\subsection{Cooling efficiency}
\label{jancooling}

Any cooling process that is due to a pressure gradient can be usefully
compared with the thermodynamic entropy-conserving (adiabatic) process,
where the same pressure is employed for cooling. Let $(p_{\rm in},
T_{\rm in})$ and $(p_{\rm out}, T_{\rm out})$ be, respectively, the
input and output pressure and temperature, and cooling $T_{\rm out}<
T_{\rm in}$ is achieved due to $p_{\rm out}< p_{\rm in}$.  For the
considered ideal-gas model, the lowest temperature $T_{\rm out, \,
  ad}$ reached adiabatically reads [see Appendix \ref{idol}]:
\begin{eqnarray}
  \label{eq:44}
    \frac{T_{\rm out,\, ad}}{T_{\rm in}}=
  \left(\frac{p_{\rm out}}{p_{\rm in}}\right)^{1/\hat{c}_p},
\end{eqnarray}
where $\hat c_p$ is defined in (\ref{eq:12}); see also Table~\ref{tab1}. 

Hence one defines the cooling efficiency \cite{silverman}:
\begin{eqnarray}
  \label{bars}
  \xi = {T_{\rm out,\,ad}}/{T_{\rm out}},
\end{eqnarray}
which has the standard meaning of efficiency (result over effort),
since the achieved result of cooling is related with $1/{T_{\rm
    out}}$.  The pressure difference is a resource and it is
quantified by $1/{T_{\rm out,\,ad}}$, hence definition (\ref{bars})
\footnote{Eq.~(\ref{bars}) is different from the coefficient of
  performance (COP) of refrigerators, which is defined via the ratio
  of the heat transferred in refrigeration over the external work
  performed to achieve this transfer. We do not need the COP, since in
  our set-ups the cooling is not achieved due to external work. }.

When quantifying cooling, people sometimes employ the Hilsch
efficiency \cite{hilsch,gutsol}:
\begin{eqnarray}
  \label{hilsch}
  \xi_{\rm
    H}= \frac{T_{\rm in}-T_{\rm out}}{T_{\rm in}-T_{\rm out,\,ad}}.
\end{eqnarray}
The meaning of $\xi_{\rm H}$ differs from that of $\xi$, because
$\xi_{\rm H}$ directly accounts for the input temperature $T_{\rm
  in}$. But they are related. As shown by(\ref{bars}), for $T_{\rm
  in}-T_{\rm out,\,ad}>0$ (a natural condition for cooling), $\xi<1$
($\xi>1$) implies $\xi_{\rm H}<1$ ($\xi_{\rm H}>1$). However,
$\xi_{\rm H}$ is less fundamental than $\xi$, since it does not appear
directly in the efficiency bound imposed by the second law. We discuss
this bound now.


\subsection{Second law bound for cooling efficiency}

The entropy balance of the fluid reads \cite{landau}
\begin{eqnarray}
  \label{eq:26_1}
&&  \partial_t(\rho s)=-\vnabla [s\rho \v-\frac{\lambda}{T}\vnabla
  T]+s_{\rm prod},
\end{eqnarray}
where $\rho s$ is the entropy density, $s\rho \v$ and
$-\frac{\lambda}{T}\vnabla T$ are, respectively, advective and thermal
entropy flux. The entropy production $s_{\rm prod}>0$ is positive due
to viscosity and heat conduction \footnote{The entropy production
  reads \cite{landau}: $s_{\rm prod} =\frac{\eta}{2T}[ \frac{\partial
    v_j }{\partial x_k} +\frac{\partial v_k }{\partial
    x_j}-\frac{2\delta_{jk} }{3} \vnabla\v ]^2 +\frac{\zeta}{T}
  [\vnabla\v]^2+\frac{\lambda [\vnabla T]^2}{T^2}>0$.}.  In the
stationary situation $\partial_t(\rho s)=0$, and (\ref{eq:26_1})
reads:
\begin{eqnarray}
  \label{eq:40}
  \frac{\d}{\d r}\left (cs-\frac{\lambda r}{T}\,\frac{\d T}{\d r}\,\right)
=r s_{\rm  prod}.
\end{eqnarray}
where we used (\ref{eq:4}). The ideal-gas entropy is
[cf. Appendix \ref{idol}]:
\begin{eqnarray}
  \label{eq:11}
  s=c_p\left(
    -  \frac{1}{\hat c_p}\ln [p]+\ln [T] -\ln\left[
      \frac{\mu}{R}  \right] \right),
\end{eqnarray}
where we employed (\ref{ideal}--\ref{ole}). 

We consider two particular cases of the adiabatic process
(\ref{eq:44}):
\begin{eqnarray}
  \label{eq:43}
  r_{\rm in}=r_2, \quad r_{\rm out}=r_1, \quad c<0, \\
  r_{\rm in}=r_1, \quad r_{\rm out}=r_2, \quad c>0.
  \label{eq:433}
\end{eqnarray}
Now (\ref{bars}, \ref{eq:40}, \ref{eq:43}, \ref{eq:433}) imply:
\begin{eqnarray}
  \label{oriogen}
  s(r_2)-  s(r_1)={\rm sign}[-c]\, c_p\ln [\xi].
\end{eqnarray}
Integrating (\ref{eq:40}) over $r$ for $r_1<r<r_2$, using $s_{\rm
  prod}>0$, (\ref{eq:11}), (\ref{eq:44}) and (\ref{oriogen}), we get
from (\ref{eq:43}, \ref{eq:433}) an upper bound for the efficiency
(\ref{bars}) that applies to both (\ref{eq:43}) and
(\ref{eq:433}):
\begin{eqnarray}
  \label{eq:45}
  |b|\ln [\xi]\leq \frac{r_1}{T(r_1)}\frac{\d T(r_1)}{\d r}
-\frac{r_2}{T(r_2)}\frac{\d T(r_2)}{\d r},
\end{eqnarray}
where $|b|$ is the Peclet number; see (\ref{eq:34.0}) and
Table~\ref{tab1}. The right-hand-side of (\ref{eq:45}) is non-zero due
to heat conductivity.  Hence if the heat-conduction is neglected
(i.e. $\lambda=0$) the cooling efficiency holds $\xi<1$; see
(\ref{eq:45}). We stress again that the inequality in (\ref{eq:45}) is
due to positivity of the entropy production: $s_{\rm prod}>0$.

\section{Boundary conditions for cooling and for permeable
  walls}
\label{definition}

When studying cooling due to a confined gasodynamic flow one should
exclude physically uninteresting cases, where the fluid is cooled due
to cold thermal baths attached to boundaries or due to external work
done by external forces.

Cooling demands that the temperature of the (radially) incoming fluid
is larger than the temperature of the outgoing fluid.  If there is a
low-temperature boundary bath, the flow should be thermally isolated
from it (adiabatic cooling). Whenever the radial flow is absent,
thermal isolation is ensured by imposing vanishing heat flux at
boundaries, e.g. at the outer boundary:
\begin{eqnarray}
  \label{eq:19}
\frac{\d T(r_2)}{\d r}=0.
\end{eqnarray}
If there is a flow through boundaries (i.e. permeable or porous
walls), (\ref{eq:19}) does not hold, because there is a heat
conductivity due to the fluid at the boundary. 

In addition to known conditions for continuity of temperature and heat
flux \cite{landau}, there is now a specific condition to be satisfied
on the adiabatic, permeable surface. To understand the origin of this
condition, let us ``decompose'' the macroscopically homogeneous,
permeable adiabatic outer surface into holes and solid parts. Recall
that $(r,\phi,z)$ are the cylindric coordinates.  Now for $(\phi,z)\in
{\rm hole}$ and a small positive $\delta$, we get that $\frac{\d}{\d
  r}T(r_2-\delta; \phi,z)$ stays finite for $\delta\to 0+$. For
$(\phi,z)\in {\rm solid}$, $\frac{\d}{\d r}T(r_2-\delta; \phi,z)$ goes
to zero for $\delta\to 0$. Hence $|\frac{\d}{\d
  r}T(r_2-\delta)|>|\frac{\d}{\d r}T(r_2)|$ after averaging over
$(\phi,z)$ that recovers the macroscopically homogeneous permeable
wall. Hence instead of (\ref{eq:19}) we obtain the following boundary
condition
\begin{gather}
  \label{eq:99}
  \left|\frac{\d T(r_2-\delta)}{\d r}\right|> \left|\frac{\d
      T(r_2)}{\d r} \right|, ~~ {\rm
    sign}\left[\frac{\d T(r_2)}{\d r}\right]\, \frac{\d^2 T(r_2)}{\d r^2}< 0,
\end{gather}
where ${\rm sign}[a]=1$ if $a\geq 0$ and ${\rm sign}[a]=-1$ if $a<0$,
and where the second inequality in (\ref{eq:99}) follows from the
first one under $\frac{\d}{\d r}T(r_2)\not =0$ and $\delta\to
0+$. Naturally, the first inequality in (\ref{eq:99}) also holds for
$\frac{\d}{\d r}T(r_2)=0$, i.e. for an adiabatic wall.
 
Likewise, we have for the thermally isolated inner wall (for
$\delta\to 0+$):
\begin{gather}
  \label{eq:999}
  \left|\frac{\d T(r_1+\delta)}{\d r}\right|> \left|\frac{\d
      T(r_1)}{\d r} \right|, ~~ {\rm
    sign}\left[\frac{\d T(r_1)}{\d r}\right]\, \frac{\d^2 T(r_1)}{\d r^2}>0,
\end{gather}

We stress that in the present model (with or without the radial flow)
it is trivial to get {\it arbitrary} low temperatures in between of
two cylinders. But generally these temperature profiles do not hold
the boundary conditions (\ref{eq:99}) or (\ref{eq:999}), i.e.  such
scenarios of low temperatures do not constitute proper cooling, since
they require that low temperatures {\it pre-exist} via boundary
baths. In particular, Appendix \ref{couette} works out the Couette
flow (laminar flow between 2 infinite rotating cylinders without
radial motion) showing that the inhomogeneous temperature profile
generated in this flow does not constitute cooling.

Conditions similar to (\ref{eq:99}, \ref{eq:999}) are deduced for the
radial velocity $v_r$ on a partially permeable wall. This is similar
to the previous case in that $v_r=0$ for an impermeable wall; see
(\ref{eq:19}). A derivation analogous to that of (\ref{eq:99},
\ref{eq:999}) produces [cf.~(\ref{eq:4})]
\begin{eqnarray}
  \label{peppy}
{\rm sign}[c]\,  \frac{\d v_r(r_2)}{\d r}<0, \\
   \label{peppyk}
{\rm sign}[c]\,  \frac{\d v_r(r_1)}{\d r}>0,
\end{eqnarray}
for the outer and inner wall, respectively. 

In the present model there are no solutions that support conditions
(\ref{eq:99}, \ref{eq:999}) and (\ref{peppy}, \ref{peppyk}) for both
inner and outer permeable walls. Thus we should put them on the wall
from which the cold flow is coming out (to ensure that low
temperatures do not exist before cooling), and left the other boundary
as a control surface assuming that both the velocity and temperature
on this surface are given.

\section{Cooling of inward flow}
\label{inward}

\subsection{Temperature profile for a weak radial flow }
\label{weak}

Recall from (\ref{eq:34}) and (\ref{eq:6.1}) that $\kappa$ and $w(x)$
are different dimensionless quantities although both are non-zero due
to radial flow. We assume that $w(x)$ and its derivatives are
small. Hence factors $(\frac{\kappa}{2}+2)\frac{w^2}{x^2}$ and
$(\chi+\frac{4}{3}) \frac{ww'}{x}$ are neglected in (\ref{s2}). This
can be done provided that $x$ is not very small.  But the influence of
the radial flow on the vortex characteristics is not neglected,
i.e. $\kappa\not =0$; see (\ref{eq:6.2}--\ref{eq:6.1}). Now the
remainder of (\ref{s2}), i.e. $(\frac{\kappa}{2}+1) \hat{v}_\phi^2 - x
\hat{v}_\phi\hat{v}'_\phi +b\hat{T}-x\hat{T}' -\beta=0$ can be solved
explicitly as
\begin{eqnarray}
  \hat{T}(x)=g(x)+\frac{\beta}{b}+x^b\, C,
\label{katon}
\end{eqnarray}
where $C$ is a constant, and where 
\begin{gather}
  g(x)\equiv \frac{2x^\kappa
  \alpha(\alpha-1)}{b-\kappa}-\frac{(1-\alpha)^2(4+\kappa)}
{2(2+b)x^2}
-\frac{\kappa\alpha^2 x^{2+2\kappa}}{2(2-b+2\kappa)}.
\label{gomes}
\end{gather}
Now $\beta$ and $C$ in (\ref{katon}) are conveniently expressed via
$\hat{T}(1)$ and $\hat{T}'(1)$, and the temperature profile reads from
(\ref{katon}):
\begin{gather}
  \label{lopes}
  \hat{T}(x) - \hat{T}(1) =\frac{x^b-1}{b}\,[\, \hat{T}'(1)-g'(1)
\,]+g(x)-g(1),
\end{gather}
The approximation that led to (\ref{lopes}, \ref{gomes}) is confirmed
by solving numerically full equations (\ref{s2}, \ref{s4}); see
Figs.~\ref{f1} and \ref{f3}.

So far the weak radial flow approximation amounted to neglecting the
radial velocity $v_r$ in the energy equation (\ref{s2}), but retaining
it in the angular Navier-Stokes equation (\ref{eq:3}); see also
(\ref{eq:6.2}--\ref{eq:6.1}). If we neglect the radial velocity $v_r$
{\it also} in the radial Navier-Stokes equation (\ref{eq:1}) [or
equivalently in (\ref{s4})] we obtain
\begin{eqnarray}
  \label{eq:14}
  \rho(r) {v_\phi^2(r)}/{r}=\d p/\d r.
\end{eqnarray}
This known equation is solved as \footnote{Let us mention the simplest
  (but incorrect) argument for the Ranque effect.  Setting
  $\rho(r)$=constant in (\ref{eq:14}) and using the ideal gas law
  $T(r)\propto p(r)$, shows that $T(r)$ is an increasing function of
  $r$ (i.e. a radial temperature separation is achieved), formally
  resembling the Ranque effect. The problem with this argument is that
  imposes a constant $\rho$. This may look formally consistent with
  other equations, but it is incorrect, e.g. because it applies also
  for $v_r=0$ (no radial motion whatsoever), while our detailed
  analysis of this $v_r=0$ situation shows that no cooling scenarios
  are possible, because the proper boundary conditions are not
  satisfied; see Appendix \ref{couette}. }
\begin{eqnarray}
  \label{eq:27}
\frac{p(x)}{p(1)}=
\exp\left[-\frac{\kappa \hat{c}_p}{b}
\int_x^1 \frac{\d y\,\hat{v}_\phi^2(y)}{y\hat{T}(y)}
\right],
\end{eqnarray}
where $\hat{T}(x)$ and $\hat{v}_\phi(x)$ are given by (\ref{lopes},
\ref{gomes}) and (\ref{eq:6}, \ref{eq:6.0}), respectively.
Eq.~(\ref{eq:27}) shows that the (dimensionless) pressure is a
monotonically increasing function of $x$.

There are cases, where (\ref{lopes}, \ref{gomes}) are valid, but
(\ref{eq:14}) [and (\ref{eq:27})] is not.  In section \ref{semi} we
show an important example of this type, where even if $w(x)\to 0$ is
imposed in the vicinity of $x=1$, it does not hold for $x<1$, because
$w(x)$ grows fast.

\subsection{Boundary conditions for inward radial flow}
\label{isothermal}

For inward flow $c\leq 0$ (hence $b\leq 0$ and $\kappa\leq 0$) we
study cooling scenarios, where the higher temperature fluid enters
into the system through the outer boundary at $x\equiv r/r_2=1$. Now
for adiabatic boundary conditions, the lower temperature fluid leaves
the system through the inner thermally isolated boundary at $x=x_0<1$
[cf.~(\ref{eq:999})]:
\begin{eqnarray}
  \label{ada}
  \hat{T}(1)>\hat{T}(x_0), \qquad \hat{T}''(x_0)>0.
\end{eqnarray}
No specific conditions are put at $x=1$, i.e. it is taken as a control
surface.

The adiabatic boundary condition (\ref{ada}) relates to
the isothermal situation [$x_0\leq x\leq 1$]
\begin{eqnarray}
\hat{T}(1)=\hat{T}(x_0)>\hat{T}(x),
\label{defo}
\end{eqnarray}
where $\hat{T}(x)$ assumes a minimum at some $x=x_{\rm min}$.
Whenever (\ref{defo}) holds, one can take $x_0\gtrsim x_{\rm
  min}$ and this produce an example of (\ref{ada}).

Eq.~(\ref{defo}) does not refer to a practically useful situation,
since no cold fluid really comes out. Nevertheless, it is interesting,
since the expected of behavior of the temperature is that it is larger
inside of the fluid, i.e.  for $\hat{T}(1)=\hat{T}(x_0)$ we expect
$\hat{T}(x)>\hat{T}(1)=\hat{T}(x_0)$ \cite{landau}. (The expectation
is also confirmed by the example of the Couette flow in Appendix
\ref{couette}.) This is because viscosity|which dissipates energy in
the bulk of the fluid|generates heat that must be transported out of
the boundaries \cite{landau}. The expected behavior holds for
$c>0$. But there are isothermal and adiabatic cooling scenarios for
$c<0$; see Figs.~\ref{f1} and \ref{f3}. We now turn to discussing
them.

\subsection{Cooling via quasi-solid vortex}
\label{quasisolid}

Let us start with a quasi-solid vortex in (\ref{eq:6},
\ref{eq:6.0}):
\begin{eqnarray}
  \label{boyard}
\alpha=1 ~~{\rm or}~~  {v_1}/{v_2} =   
\left({r_1}/{r_2}\right)^{1+\kappa}.
\end{eqnarray}
The temperature for this situation reads from (\ref{lopes},
\ref{gomes}):
\begin{gather}
\hat{T}(x)-\hat{T}(1)=\frac{[\,\hat{T}'(1)
+\frac{\kappa}{2}\,][x^b-1]}{b}
+\frac{\kappa (x^b-x^{2+2\kappa})}{2(2+2\kappa-b)}.
\label{boris}
\end{gather}
Let us see to which extent (\ref{boris}) can hold 
condition (\ref{defo}). 
Now $\hat{T}'({x}_{\rm min})=0$ leads to
\begin{eqnarray}
\label{godunov}
{x}_{\rm min}^{2-b+2\kappa}=1+\frac{(2-b+2\kappa)\hat{T}'(1)}{\kappa(1+\kappa)},
\\
\hat{T}''({x}_{\rm min})=-\kappa(1+\kappa){x}_{\rm min}^{2\kappa}.
\label{yang}
\end{eqnarray}
Eq.~(\ref{godunov}) means that $\hat{T}'({x}_{\rm min})=0$ has only
one solution. Since this solution ought to be a minimum
[cf.~(\ref{defo})], we have to require $\hat{T}'(1)>0$ that together
with $0\leq x\equiv\frac{r_1}{r_2}\leq 1$ leads from (\ref{godunov},
\ref{yang}) to $\kappa(1+\kappa)<0$ and $2(1+\kappa)>b$, or
equivalently to
\begin{eqnarray}
  \label{yard}
 -1< \kappa <0, \quad
 0<\hat{T}'(1)<-\frac{\kappa(1+\kappa)}{2(1+\kappa)-b}. 
\end{eqnarray}
Thus under conditions (\ref{yard})|and naturally $x_0$ sufficiently
smaller than $x_{\rm min}$|we get an isothermal cooling scenario
(\ref{defo}). Taking $x_0\gtrsim x_{\rm min}$ we get instead
an example of the adiabatic scenario (\ref{ada}); cf. the discussion
after (\ref{defo}).

Eq.~(\ref{boyard}, \ref{yard}) imply a quasi-solid vortex that is
frequently observed experimentally. Examples of the above cooling
scenario are presented in Figs.~\ref{f1} and \ref{f3} for isothermal
and adiabatic scenarios, respectively. Naturally, the cooling takes
place both in terms of (thermodynamic) temperature $\hat{T}$ and
stagnation enthalpy $\hat{U}$. We shall see below that conditions
(\ref{boyard}, \ref{yard}) are sufficiently representative, i.e.
more general cooling scenarios implied from (\ref{lopes})
are close to those predicted by (\ref{boyard}, \ref{yard}).

\subsection{Magnitude and efficiency  of cooling}

\begin{figure*}
\includegraphics[width=12.6cm]{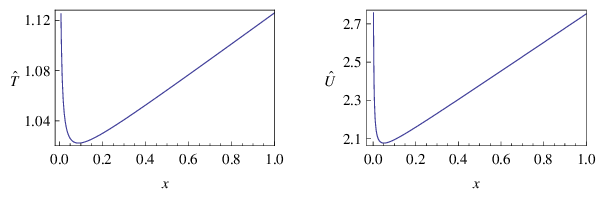}
\caption{Isothermal cooling with inward flow ($c<0$) and
  quasi-solid vortex; see section \ref{quasisolid}. Dimensionless temperature $\hat{T}(x)$ and 
dimensionless stagnation enthalpy $\hat{U}(x)$ versus dimensionless distance $x$. \\
  Left (right) figure: $\hat{T}(x)$ ($\hat{U}(x)$) versus $x=r/r_2$ for $x\in [x_0,1]$, where $x_0=0.007347$. \\
  The curves are obtained from numerical solution of (\ref{s2},
  \ref{s4}) for $\alpha=1$, $\kappa=-0.5$, $b=-1$, $\hat{c}_p=3.5$,
  $\beta=-1$ ($\hat{T}'(1)=0.124$), $\chi=10$ and $\hat{T}(1)=1.126$,
  $w(1)=10^{-4}$, $w'(1)=0$. These parameter values are consistent
  with experiments \cite{savino}; see (\ref{star}), (\ref{yard}) and
  the related discussion, as well as
  (\ref{eq:63}), (\ref{eq:66}) and the related discussion. \\
  The minimal dimensionless temperature
  is $\hat{T}_{\rm min}=\hat{T}(x_{\rm min})=1.0223$, $x_{\rm min}=0.008891$.\\
  The energy values are: $\Delta \widehat{E}=-0.24699$,
  $\Delta\widehat{W}=0.48228$, $\Delta\widehat{W}_r=-0.01405$,
  $\widehat{Q}(x_0)=-x_0\hat{T}'(x_0)=0.11129$,
  $\widehat{Q}(1)=-\hat{T}(1)=-0.124$; cf.~(\ref{energy}--\ref{1law}). \\
  The pressure $p(x)$ is a monotonically increasing function of
  $x\in [x_0,1]$; cf.~(\ref{eq:27}). $\hat{T}_{\rm ad}(x_{\rm
    min})=0.73556$. The Hilsch efficiency is $\xi_{\rm H}=0.26559$; cf.~(\ref{hilsch}). \\
  Upon decreasing the initial temperature, $x_{\rm min}$ increases,
  while both the magnitude and quality of cooling decrease, e.g. for
  $\hat{T}(1)=1.16$ we get $x_{\rm min}=0.50597$, $\hat{T}_{\rm
    min}=\hat{T}(x_{\rm min})=1.13083$ and $\xi_{\rm H}=0.13016$.\\
  $w(x)$ (and hence $w(x)/x$) is a monotonically decreasing function
  of $x\in [x_0,1]$. Thus condition (\ref{peppyk}) holds for the inner
  wall. Condition (\ref{peppy}) for the outer wall cannot hold
  simultaneously with (\ref{peppyk}); this is why a permeable wall is
  not imposed at $r=r_2$. The latter is just a control surface within
  which we describe the flow.}
\label{f1}
\end{figure*}

\begin{figure*}
\includegraphics[width=11.cm]{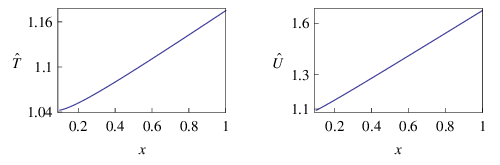}
\caption{Adiabatic cooling with inward flow ($c<0$) and quasi-solid vortex. 
Dimensionless temperature $\hat{T}(x)$ and dimensionless stagnation 
enthalpy $\hat{U}(x)$  versus dimensionless distance $x$. \\
  Left (right) figure: $\hat{T}(x)$
  ($\hat{U}(x)$) versus $x=r/r_2$ for $x\in [x_0,1]$, where $x_0=0.1$. \\
  The curves are obtained from numerical solution of (\ref{s2},
  \ref{s4}) for $\alpha=1$, $\kappa=-0.5$, $b=-0.5$, $\hat{c}_p=3.5$,
  $\beta=-0.5$, $\chi=10$ and $\hat{T}(1)=1.175$,
  $w(1)=10^{-4}$, $w'(1)=0$.\\
  The minimal temperature is $\hat{T}(x_0)=1.04233$. As required for
  an adiabatic, permeable inner boundary: $\hat{T}''(x_0)=1.48034$; 
  cf.~(\ref{eq:99}, \ref{eq:999}). \\
  The energy values are: $\Delta \widehat{E}=-0.291278$,
  $\Delta\widehat{W}=0.44866$, $\widehat{Q}(x_0) =-x_0\hat{T}'(x_0)=-0.0051$,
  $\widehat{Q}(1)=-\hat{T}'(1)=-0.1625$; cf.~(\ref{energy}--\ref{1law}). \\
  Efficiency: $\hat{T}_{\rm ad}(x_0)=0.52012$; see (\ref{eq:44},
  \ref{eq:43}). The Hilsch efficiency is $\xi_{\rm
    H}=0.2026$. Pressure $p(x)$ and $w(x)$ hold $p'(x)>0$ and
  $w'(x)<0$ for $x\in [x_0,1]$.  }
\label{f3}
\end{figure*}

\begin{figure*}
\includegraphics[width=12.4cm]{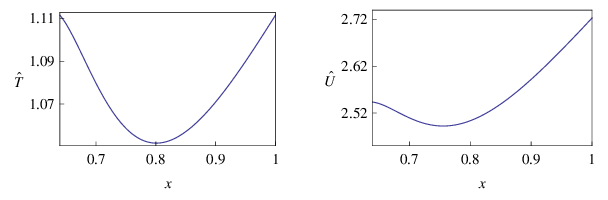}
\caption{ Isothermal cooling with inward flow and $\alpha=-0.5$ vortex. 
Dimensionless temperature $\hat{T}(x)$ and dimensionless stagnation 
enthalpy $\hat{U}(x)$  versus dimensionless distance $x$. \\
  Left (right) figure: $\hat{T}(x)$
  ($\hat{U}(x)$) versus $x=r/r_2$ for $x\in [x_0,1]$, where $x_0=0.639472$. \\
  The curves are obtained from numerical solution of
  (\ref{s2}, \ref{s4}) for $\alpha=-0.5$, $\kappa=-4$, $b=-5$,
  $\hat{c}_p=3.5$, $\beta=-7$, $\chi=10$
  and $\hat{T}(1)=1.1115$, $w(1)=0.01$, $w'(1)=0$. \\
  The minimal dimensionless temperature
  is $\hat{T}_{\rm min}=1.05166$ and it is reached for $x_{\rm min}=0.80077$.\\
  The energy values are: $\Delta E=-1.62148$, $\Delta W=2.26812$, $\Delta
  W_r=-0.01403$, $\widehat{Q}(x_0)=-x_0\hat{T}'(x_0)=0.204143$,
  $\widehat{Q}(1)=-\hat{T}'(1)=-0.4425$. \\
  The pressure $\hat{p}(x)$ ($w(x)$) is a monotonically increasing
  (decreasing) functions of $x\in [x_0,1]$. $\hat{T}_{\rm
    ad}(x_0)=0.952581$; see (\ref{eq:44}, \ref{eq:43}). The Hilsch
  efficiency is $\xi_{\rm H}=0.376542$; cf.~(\ref{hilsch}).  }
\label{f4}
\end{figure*}

Both adiabatic and isothermal cooling scenarios lead to 
relatively weak effects in the sense of
\begin{eqnarray}
  \label{eq:18}
  \frac{\hat{T}(1)-\hat{T}(x_{\rm min})}{{\rm min}[1,\hat{T}(1)]}
\simeq 0.01 - 0.1.
\end{eqnarray}
The cooling magnitude in terms of stagnation enthalpy is larger; see
Figs.~\ref{f1}-\ref{f4}.

The efficiency (\ref{bars}) of cooling under condition (\ref{ada})
is smaller than $1$:
\begin{eqnarray}
  \label{eq:2}
  \xi<1.
\end{eqnarray}
Whenever $\frac{\d}{\d r} T(r_1)$ is sufficiently small, (\ref{eq:2})
follows directly from the second law bound (\ref{eq:45}), where
$\frac{\d}{\d r} T(r_2)>0$; cf.~(\ref{eq:43}). Otherwise, (\ref{eq:2})
is confirmed numerically; see Figs.~\ref{f1}--\ref{f4}. Hence the
Hilsch efficiency (\ref{hilsch}) also holds $\xi_{\rm H}<1$, as shown
by Figs.~\ref{f1}--\ref{f4}.

For both isothermal and adiabatic cooling scenarios we obtain from
(\ref{oriogen}, \ref{eq:2}) for the entropy difference:
\begin{eqnarray}
  \label{stokes}
s(r_2)-s(r_1)=c_p\ln[\xi]<0.
\end{eqnarray}
Hence the final entropy is always larger than the
initial one: ${s}(r_1)>{s}(r_2)$. 


\subsection{Work and energetics}

As shown by (\ref{vagr}, \ref{boyard}), the work done by rotating
cylinders is positive under conditions (\ref{yard}):
\begin{eqnarray}
  \label{eq:24}
\Delta  \widehat{W}_\phi=-\kappa(1-x_0^{2+2\kappa})>0,
\end{eqnarray}
which means that the work is extracted. The work $\widehat{W}_r$ done
by radial external forces is small, but negative (i.e. it is
invested), and the overall work is positive; see Figs.~\ref{f1} and
\ref{f3}. Thus the set-up does not demand an external investment of
work \footnote{We mention another scenario of cooling, which is
  realized for the inward flow and the potential vortex
  $\hat{v}_\phi(x)=1/x$; see (\ref{eq:6.0}) with $\alpha=0$. This
  scenario is less interesting, since it is driven by external
  investment of work [$W_\phi'<0$, as seen from (\ref{kon},
  \ref{eq:32})], while its efficiency and magnitude hold the same
  constraints (\ref{eq:2}, \ref{eq:18}). An interesting point of this
  scenario is that it is accompanied by the kinetic energy that
  increases in the direction of the flow: $\left[
    \frac{|\kappa|}{2}\hat{v}_\phi^2 +|b|\hat{T}(x) \right]'\leq 0$ in
  (\ref{eq:38}). Appendix \ref{potential_vortex} studies details of
  this scenario.}: cooling takes place due to the initial pressure
larger than the final one, $p(1)>p(x_0)$. In other words, cooling
takes place due to the initial potential energy of the fluid.

Under isothermal boundary conditions both thermal baths (at $x=1$ and
$x=x_0$, respectively) provide heat to the system.  Using (\ref{defo},
\ref{boyard}) (and the fact that $w(x)$ is assumed to be small), we
get from (\ref{energy}, \ref{1law})
\begin{eqnarray}
\label{maga}
\Delta \widehat{E}=\Delta \widehat{E}_{\rm kin}
=\frac{\kappa}{2} (1-\hat{v}_\phi^2(x_0)),
\end{eqnarray}
which is negative due to (\ref{boyard}). Hence we also get cooling in
terms of the stagnation enthalpy; see (\ref{stop}, \ref{energy}).
Eqs.~(\ref{eq:24}, \ref{maga}) are consistent with the first law
(\ref{1law}), which for the present situation reads: $\Delta
\widehat{E}+\Delta\widehat{W}=|\widehat{Q}_1|+|\widehat{Q}_2|$.

These features hold for other cases of isothermal and adiabatic
cooling.  Fig.~\ref{f4} shows an isothermal scenario with
$\alpha=-0.5$, where $\hat{v}_\phi(x)$ is again a concave, increasing
function of $x$.  Fig.~\ref{f3} demonstrates an adiabatic cooling
scenario that does not reduce to the isothermal case (whenever the
latter holds one can obtain an adiabatic scenario by taking
$x_0>x_{\rm min}$).


Let us write from (\ref{bora}, \ref{energy}--\ref{heat})
\begin{eqnarray}
  \label{eq:38}
  0= \left[ -\widehat{E}(x)+x\hat{T}'(x)
-\widehat{W}_\phi(x)-\widehat{W}_r(x)\right]'.
\end{eqnarray}
In (\ref{eq:38}) we assume that isothermal boundary conditions hold
for $c<0$; hence $b$ and $\kappa$ are negative. Now
$[-\widehat{E}(x)]'>0$ for $x_{\rm min}\lesssim x$, because this means
cooling in terms of the stagnation enthalpy. One also has
$[ x\hat{T}'(x)]'>0$ for $x_{\rm min}\approx x$. It appears
also that $[-\widehat{W}_r(x)]'>0$ has the same sign as
$[-\widehat{E}(x)]'>0$. Moreover, it quickly prevails over other
factors; this is why for $c<0$ cooling exists only in the weak radial
flow situation, where $\widehat{W}_r\to 0$.

Thus for holding (\ref{eq:38}) and achieving cooling we need
\begin{eqnarray}
  \label{kon}
\widehat{W}_\phi'(x) >0,  
\end{eqnarray}
which|using (\ref{work_tangential}, \ref{eq:6}, \ref{eq:6.0})|is
equivalent to
\begin{eqnarray}
  \label{eq:32}
  0>4(1-\alpha)^2 &+&x^{2+\kappa}\alpha (1-\alpha)\kappa(\kappa-2)\nonumber\\
  &+&2x^{4+2\kappa}\alpha^2 \kappa(\kappa+1).
\end{eqnarray}
Hence the validity of (\ref{kon}) (at least for certain values of
$x$), i.e. the positivity of work, is a necessary condition for both
isothermal and adiabatic cooling in the regime $c<0$. This condition
was obtained in \cite{penge}, but its necessary character was not
properly stressed, in particular, because the heat conductivity and
boundary conditions necessary for cooling were neglected.  In
particular, (\ref{kon}) can lead to mistakes if it is taken as a
sufficient condition for cooling.


\subsection{Dependence of temperature profiles on the radial Reynolds
  number $\kappa$} 

The radial Reynolds number $\kappa=c/\eta$ [see Table~\ref{tab1}]
combines the radial flow $c$ and the viscosity $\eta$. Hence it is
important to understand how the cooling temperature profiles depend on
$\kappa$. 

Now $\kappa$ can change|from one fluid to another|due to $c$ and/or
due to $\eta$. We shall focus on the later scenario recalling that
$\eta$ can be an effective (eddy or turbulent) viscosity; see section
\ref{relo} for details. Anticipating some of discussions from section
\ref{relo}, we recall that the effective viscosity changes together
with the effective heat-conductivity $\lambda$ such that the Prandtl
number ${\rm Pr}=b/\kappa$ is roughly a constant (also in the
turbulent regime) \cite{shirokov}. We also assume that the quasi-solid
condition (\ref{boyard}) holds, i.e. the ratio $v_1/v_2$ changes
together with $\kappa$ so that $\alpha=1$ is kept fixed. All these
conditions are observed in vortex tubes \cite{gutsol,savino}.

\begin{figure*}
\includegraphics[width=7cm]{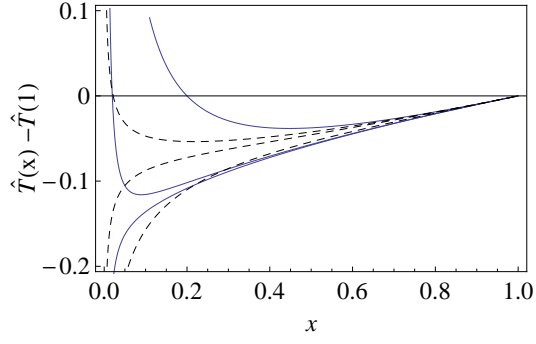}
\caption{Dependence of the dimensionless temperature $\hat{T}(x)$
  profiles [see (\ref{lopes})] on the radial Reynolds number, where
  the Prandtl number ${\rm Pr}=b/\kappa=2$ is fixed, the quasi-solid
  vortex condition $\alpha=1$ is obeyed [see (\ref{boyard})], and
  $\hat{T}'(1)=0.1$. \\ Full, blue curves from top to bottom:
  $\kappa=-0.5$, $\kappa=-0.72$ and $\kappa=-0.725$. The first curve
  holds condition (\ref{ada}) for all $x_0$, the second one holds them
  for $x_0<0.21$, while the latter curve (with $\kappa=-0.725$) is
  void of physical meaning, since neither the adiabatic boundary
  condition (\ref{ada}), nor the isothermal condition (\ref{defo}) are
  satisfied for it. \\ Dashed, black curves from top to bottom:
  $\kappa=-0.3$, $\kappa=-0.25$ and $\kappa=-0.15$. For the curve with
  $\kappa=-0.25$ the adibatic boundary condition (\ref{ada}) holds for
  $x_0>0.447$. The whole curve with $\kappa=-0.15$ is void of the
  physical meaning, since neither (\ref{ada}), nor (\ref{defo}) hold.
}
\label{f21}
\end{figure*}

Fig.~\ref{f21} shows temperature profiles (\ref{lopes}) for different
values of $\kappa$. (Recall that the approximate formula (\ref{lopes})
is well-confirmed numerically). It is seen that the cooling effect
disappears both for sufficiently small and large values of $\kappa$
(which we recall is negative for the outward flow due to $c<0$; see
Table~\ref{tab1}). This is because the boundary condition (\ref{ada})
ceases to hold. The cooling effect is locally maximal before its
disappearance; see Fig.~\ref{f21}.

\subsection{Dependence of cooling on the rotation speed}
\label{depo}

Since the angular (vortical) motion is necessary for above cooling
scenarios, we turn to studying in detail the dependence of cooling
temperature profiles on the rotation speeds $v_1$ and $v_2$ of
cylinders. It is natural to assume that these variables change by
keeping the ratio $v_1/v_2$ fixed. Now $\alpha$ is fixed parameter as
well [see (\ref{eq:6.1})] and hence we stay within the quasi-solid
vortex regime. This is useful, because this regime is observed
experimentally for a broad regime of experimental parameters
\cite{gutsol,savino}.

Since now $v_2$ is a variable we need to redefine $\hat{T}$ given by
(\ref{eq:34}); see also Table~\ref{tab1}. Instead
of $\hat{T}$, we employ $\hat{T}_0=\frac{v_2^2}{v_{2,0}^2}
\hat{T}=\frac{\lambda}{v_{2,0}^2\eta} T$ that equals to $\hat{T}$
evaluated at some fixed reference value $v_{2,0}$ of the velocity
$v_2$, i.e. $\hat{T}_0$ does not already have a parametric depedence
on $v_2$. Now we get instead of (\ref{lopes}):
\begin{align}
  \hat{T}_0(x) - &\hat{T}_0(1) =\frac{x^b-1}{b}\,[\, \hat{T}_0'(1)
-\epsilon g'(1)\,]\nonumber\\
&+\epsilon[\,g(x)-g(1)\,], \quad\qquad \epsilon\equiv {v_2^2}/{v_{2,0}^2},
  \label{lopes2}
\end{align}
where $g(x)$ is still defined by (\ref{gomes}), i.e. it does not have any
parametric dependence on $v_2$ or on $\epsilon$. According to
(\ref{lopes2}), a larger $\epsilon$ means a bigger deviation of $v_2$
from its reference values. 

\begin{figure*}
\includegraphics[width=7cm]{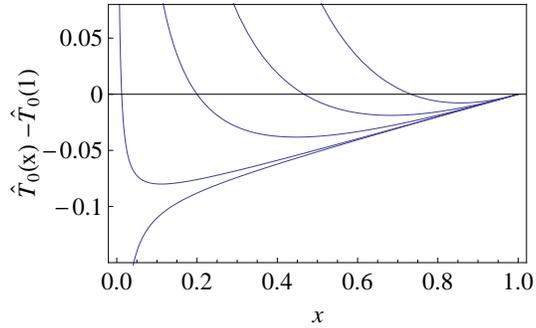}
\caption{Dependence of temperature profiles
  $\hat{T}_0(x)-\hat{T}_0(1)$ on the rotation speed according to
  (\ref{lopes2}). A larger $\epsilon$ means a bigger deviation of
  $v_2$ from its reference values. From bottom to top: $\epsilon=0.78$ (no
  cooling is present, since neither (\ref{ada}), nor (\ref{defo})
  hold), $\epsilon=0.81$, $\epsilon=1$, $\epsilon=1.5$ and
  $\epsilon=3$. The latter four curves demonstrate cooling. For other parameters we take
  [cf.~Figs.~\ref{f1} and \ref{f3}]: $\hat{T}'_0(1)=0.1$, $\alpha=1$,
  $b=-1$, $\kappa=-0.5$.  }
\label{f22}
\end{figure*}

Fig.~\ref{f22} shows temperature profiles (\ref{lopes2}) for different
values of $\epsilon$ and for parameters given by (\ref{boyard},
\ref{yard}). For given values of parameters, there is a value of
$v_2$, where the cooling effect is maximal, i.e. the lowest
temperature is reached. For the parameters of Fig.~\ref{f22} this
critical value is found from $\epsilon =0.8$. When $v_{2}$ decreases
from this critical value, the cooling effect ceases to exist, since
the cooling boundary conditions|as given by (\ref{ada}) or
(\ref{defo})|cannot hold anymore, i.e. the curve in Fig.~\ref{f22}
that corresponds to $\epsilon=0.78$ is void of physical meaning. When
$v_{2}$ increases from the critical value, the magnitude of vortex
cooling|as measured by the lowest temperature reached|monotonously
decreases. In particular, for a larger $\epsilon$, we need to take a
larger $x_0=r_1/r_2$ (i.e. $x_0\to 1$) to achieve cooling; see
Fig.~\ref{f22}.

\section{Relations with experiments}
\label{relo} 

\subsection{Effective viscosity}
\label{effo}

The actual flow in vortex tubes is highly turbulent
\cite{gutsol}. Hence if one uses hydrodynamic equations (in
particular, Navier-Stockes equations) with constant values of
viscosities $\eta$ and $\zeta$ and heat-conductivity $\lambda$, it is
at very least necessary to employ there effective (i.e. turbulent or
non-molecular) estimates for these parameters \cite{landau,shirokov}.

Taking into account that the considered flow is confined and
inhomogeneous (i.e. there are radial and angular flows) we choose to
estimate the turbulent viscosity $\eta$ via the Nusselt's formula
\cite{nusselt}, which was originally proposed for estimating the
turbulent viscosity in pipes. Ref.~\cite{shirokov} found that this
formula applies for describing compressible turbulence in a
sufficiently wide range of Reynolds numbers. The formula reads
\cite{nusselt,shirokov}:
  \begin{eqnarray}
    \label{eq:63}
    \eta=0.15 \,\eta_{\rm mol} 
    \left[\,\rho\, v_\phi\, l\,/\eta_{\rm mol}\,\right]^{3/4},     
  \end{eqnarray}
  where $v_\phi$ is the characteristic value of the angular velocity,
  and $l$ is the characteristic length, and $\eta_{\rm mol}$ is the
  molecular viscosity: $\eta_{\rm mol}=1.8\times 10^{-5}$ kg/(m\,s)
  for the air. Also, taking in (\ref{eq:63}) typical experimental
  parameters for vortex tubes \cite{savino}: $\rho=1.2 $ kg/${\rm
    m}^3$, $v_\phi\simeq v_{\rm sound}=$331 m/s and $l=0.1$ m, we get
  \begin{eqnarray}
    \label{bratva}
\eta \sim
  0.085\,\, {\rm kg/(m\, s)},
      \end{eqnarray}
      which is several orders of magnitude larger than $\eta_{\rm
        mol}$.  The estimate (\ref{bratva}) roughly coincides with an
      estimate given in \cite{knoer} via the mixing-length formula
      \cite{shirokov}: $\eta=\rho \ell^2 \frac{\d v_\phi}{\d r}$,
      where $\ell=\beta \frac{\d v_\phi}{\d r}\left /\frac{\d^2
          v_\phi}{\d r^2}\right.$ is the mixing length and where
      $\beta$ is a suitable numerical constant. It is this specific
      form of the mixing-length formula that can apply to compressible
      turbulence \cite{shirokov}.


\subsection{Comparison with experiment}
\label{compo}

Let us recall that the present model omits several physical factors
that are met in realistic vortex tubes; see section \ref{intro} for a
discussion of basic set-ups for vortex tubes. In particular, the axial
motion is neglected, and the fluid is removed radially (in contrast to
axial removal in vortex tubes). Hence the model set-up has
one (not two) output temperatures, and the whole outgoing fluid is
cooled (in contrast to the standard Ranque tube which has two output
flows and achieves temperature separation \cite{gutsol}).

Taking into account (\ref{star}), the magnitude of cooling is
predicted from (\ref{eq:18}) to be around 10 K.  Note that best vortex
tubes provide (starting from 300 K) a larger cooling of order 70-80 K
\cite{hilsch,graham,gutsol}. Though such a stronger effect is lacking
in the present model, we recall that in those cases only a part
(e.g. $\sim 20$ \% according to \cite{hilsch}) part of the overall
flow is strongly cooled, the remaining part is heated up. In the
present model the whole outgoing fluid is cooled.

The Hilsch efficiency $\xi_{\rm H}$ of cooling obtained in the present
model is of order of $0.1-0.4$; see Figs.~\ref{f1}-\ref{f4}. It also
agrees with experiments, though not for the most efficient vortex
tubes, where $\xi_{\rm H}$ can be as high $0.6-0.7$
\cite{hilsch,gutsol}.

The magnitude of the radial flow over the angular flow, expressed by
\begin{eqnarray}
  \label{eq:66}
w(1)=10^{-3}-10^{-4},  
\end{eqnarray}
also agrees with experimental measurements \cite{gutsol,xue}, though
it is to be stressed that these measurements were carried out for
sufficiently long vortex tubes, where the axial velocities (neglected
altogether in the present model) are definitely larger than the radial
velocities. The quasi-solid vortex (\ref{boyard}) for
$\hat{v}_\phi(x)$ is also seen experimentally, though the experimental
results also indicate that for $x\sim 1$, $\hat{v}_\phi(x)$ starts to
decay, i.e. the quasi-solid vortex $\hat{v}_\phi(x)\sim x^{1+\kappa}$
(for our model we took $\kappa\sim -0.5$) changes towards the
potential vortex $\hat{v}_\phi(x)\sim x^{-1}$ \cite{gutsol,xue}. This
change of $\hat{v}_\phi(x)$ is given much importance in certain
theories of vortex cooling \cite{graham,gutsol}, but is not present
here.

The input density is estimated from (\ref{eq:4}, \ref{eq:34}):
\begin{eqnarray}
  \label{eq:42}
  \rho(r_2)=\frac{|\kappa|\, \eta}{w(1)\,r_2\, v_\phi(r_2)}.
\end{eqnarray}
Estimating $r_2\simeq 0.1$ m (reasonable value for the outer radius of
a vortex tube), $v_\phi(r_2)\simeq v_{\rm sound}$ and $\eta$ for air
as $\eta \simeq 0.085$ kg/(m\, s) [see \ref{bratva}], we end up with
$\rho(r_2)\simeq\frac{10^{-3}}{w(1)}$ kg/(m\, s) in (\ref{eq:42}).
Estimating from the present model $w(1)\simeq 10^{-4}$ (and recalling
$p=R\rho T/\mu$) we get that the input pressure is few times larger
than the atmospheric pressure \cite{ranque}.

\comment{With this value of the characteristic viscosity $\eta$
  (proposed in \cite{knoer}) we already took into account that the
  actual flow in Ranque tubes is turbulent. Hence if one employs a
  laminar model for describing this flow, then the viscosity of air
  has to increase from its molecular value $\eta_{\rm mol}= 1.8\times
  10^{-5}$ kg/(m\, s) to account for turbulent (eddy) viscosity $\eta$
  \cite{landau}. Qualitatively, this is done as follows \cite{landau}:
  the ratio $\eta/ \eta_{\rm mol}$ of turbulent to molecular viscosity
  is estimated as $\sim {\rm Re}/{\rm Re}_{\rm cr}$, where ${\rm Re}$
  is the (angular) Reynolds number and ${\rm Re}_{\rm cr}$ is a
  (generically sizable) numeric coefficient. Since $\eta/\eta_{\rm
    mol}\gg 1$, the Reynolds number is to defined via the turbulent
  viscosity: ${\rm Re}=\rho v l/\eta$, where $\rho$, $v$ and $l$ are
  (respectively) the characteristic values of mass density, velocity
  and linear dimension. Hence $\eta=\sqrt{\rho v l\eta_{\rm mol}/{\rm
      Re}_{\rm cr}}$. Roughly estimating as ${\rm Re}_{\rm cr}\sim
100$, and $\rho=1.2 $ kg/${\rm m}^3$, $v\simeq v_{\rm sound}=$331 m/s
and $l=0.1$ m, we get $\eta \sim 10^{-3}$ kg/(m\, s) \cite{knoer}}

Altogether, given limitations of the present model, and complications
of the flow in real vortex tubes, one can say that the model is in a
fair qualitative agreement with experiments, though it is far from
predicting (and explaining) the features of best vortex tubes, those
providing the largest efficiency or the largest magnitude of cooling.

Savino and Ragsdal presented a simplified set-up of vortex cooling
effect \cite{savino} that in several respects is similar to the
present model. They studied two short (compared to the diameters)
concentric cylinders; the length to diameter ratio was $0.1$ and $0.5$
for two different samples. (For traditional Ranque-Hilsch tubes the
length to diameter ratio is 20--50). The rotating air enters radially
from the whole outer permeable cylinder and leaves through the inner
(smaller) cylinder.  Rotational flow was created via the outer
cylinder with Mach number $\simeq 0.2$. The velocity of this flow
was much larger than that of the radial flow.  The authors found
a cooling effect in terms of the radial variation of the stagnation
enthalpy \footnote{Recall that (due to Bernoulli's theorem)
  cooling in terms of the stagnation enthalpy cannot be explained via
  adiabatic fluid dynamics. } (no data on
thermodynamic temperature or velocities was given). The magnitude of
this cooling effect is lower than predictions of the present model;
cf. the stagnation enthalpy data in Figs.~\ref{f1}--\ref{f3}. They
confirmed that the radial distribution of the stagnation enthalpy is
established already near the end-wall of the tube and is not affected
by the weak axial flow. In particular, the axial change of the
stagnation enthalpy was much smaller than the radial one. (Hence it
was legitimate to neglect the axial flow in the model.) They found
that the experimental data can be described by (\ref{eq:14}), where
pressure is balanced by the centrifugal force (this equation does not
contain the viscosity explicitly). It was observed that the pressure
decreases monotonically with the radius, as confirmed by (\ref{eq:27})
of the present model.

\section{Cooling of outward flow}
\label{semi}

\subsection{Conditions for cooling}

Now we assume that $c>0$ (i.e. $\kappa>0$ and $b>0$) and the outer
boundary of the system is thermally isolated in the sense of
(\ref{eq:99}).  Hence the outgoing fluid being colder than in-coming
one, this implies (for $c>0$) $\hat{T}'(1)< 0$, and then (\ref{eq:99})
demands
\begin{eqnarray}
  \label{eq:31}
\hat{T}''(1)>0,  
\end{eqnarray}
at the adiabatic outer boundary. No specific conditions are imposed at
the inner boundary $r=r_1$ that can be thus considered as a control
surface. 

The full expression for $\hat{T}''(1)$ is worked out from (\ref{s2},
\ref{s4}) [recall (\ref{eq:15})]:
\begin{gather}
\label{ona}
 \hat{T}''(1)= \frac{b\hat{T}'(1)w'(1)}{\hat{c}_p
  w(1)}
+[\frac{b(\hat{c}_p-1)}{\hat{c}_p}-1]\hat{T}'(1)\\
-(\chi+\frac{1}{3})
  w'(1)^2-(2w(1)-w'(1))^2-(\alpha(2+\kappa)-2)^2.
\label{on}
\end{gather}
The first line (\ref{ona}) contains potentially positive terms, while
all terms in (\ref{on}) are non-positive. Hence (\ref{eq:31}) demands
that (\ref{on}) is sufficiently small, e.g. via $w'(1)\to
0$. Likewise, $\hat{T}'(1)$ cannot be very small.

If $w(x)$ and $w'(x)$ are sufficiently small, $\hat{T}(x)$ can be
approximately determined from (\ref{lopes}, \ref{gomes}). However,
(\ref{eq:14}, \ref{eq:27}) do not apply anymore, because the
(dimensionless) pressure $\hat{p}(x)=\hat{T}(x)/w(x)$ is now a
decreasing function of $x$ for $x\in [x_0,1]$. Thus, $w(x)$ is now
essential in (\ref{s4}) and it is important for determining the
energetics. The physical reason for this is that the work
$\Delta\widehat{W}_r$ [see (\ref{work_radial})] done by viscous radial
forces is relevant, as seen below.

Fig.~\ref{f6} demonstrates the main outward-flow cooling scenario for
$\alpha=1$; cf.~(\ref{eq:6.0}). The magnitude of cooling is now
sizable
\begin{eqnarray}
  \label{eq:188}
  [\hat{T}(x_0)-\hat{T}(1)]/\hat{T}(x_0)\geq 10.
\end{eqnarray}
The temperature profile $\hat{T}(x)$ shown in Fig.~\ref{f6} 
coincides with that found from (\ref{boris},
\ref{godunov}), where now $\hat{T}'(1)<0$ and $x_{\rm min}$ in
(\ref{godunov}) should be changed to $x_{\rm max}$, because this is
now the maximum of $\hat T(x)$. Then one should take $x_0>x_{\rm max}$
so that the temperature decreases for $x_0<x <1$.

\subsection{Energetics, entropy and efficiency}

\begin{figure*}
\vspace{0.25cm}
\includegraphics[width=18.4cm]{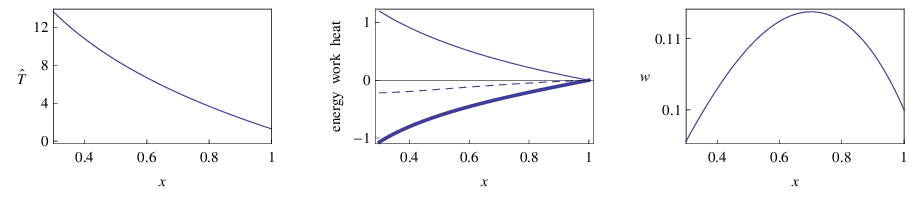}
\caption{ Adiabatic cooling with weak outward radial flow and
  quasi-solid vortex. Dimensionless temperature $\hat{T}(x)$, 
  energy quantities and dimensionless radial velocity $w$ 
  versus dimensionless distance $x$; see Table~\ref{tab1}. \\
  Eqs.~(\ref{s2}, \ref{s4}) are solved numerically for $\alpha=1$,
  $\kappa=0.1$, $b=0.1$, $\hat{c}_p=3.5$, $\beta=11$
  ($\hat{T}'(1)=-10.7867$), $\chi=10$, $\hat{T}(1)=1.295$, $w(1)=0.1$,
  and $w'(1)=-0.1$ in the range
  $x=r/r_2 \in [x_0,1]$, where $x_0=0.3$.\\
  Left figure: $\hat{T}(x)$ versus $x=r/r_2$
  (the dimensionless stagnation enthalpy $\hat{U}(x)$ behaves
  similarly). A nearly identical temperature profile is obtained upon
  solving (\ref{s2}) with $w(x)=w(1)$ and $w'(x)=0$.    \\
  Middle figure: the dimensionless energy
  $\widehat{E}(x)-\widehat{E}(1)$ (solid line), radial work
  $\widehat{W}_r(x)-\widehat{W}_r(1)$ (dashed line) and heat
  $\widehat{Q}(x)-\widehat{Q}(1)$ (bold line) versus $x$. \\
  Right figure: $w(x)$ versus $x$. Condition (\ref{peppy}) holds. It is
  written as $w(x_0)>x_0w'(x_0)$ and $w(1)>w'(1)$. The radial
  velocity $\propto w(x)/x$ is a monotonically decreasing function of $x$.\\
  The minimal (maximal) dimensionless temperature
  is $\hat{T}(1)=1.295$ ($\hat{T}(x_0)=13.6389$) and $\hat{T}''(1)=9.98$.\\
  Energetics: $\Delta \widehat{E}=-1.1925$, $\Delta \widehat{E}_{\rm
    kin}=0.04188$, $\Delta\widehat{W}_r=0.21743$,
  $\Delta\widehat{W}=0.1245$,
  $\widehat{Q}(x_0)=-x_0\hat{T}'(x_0)=9.7187$,
  $\widehat{Q}(1)=-\hat{T}'(1)=10.885$. \\
  Efficiency: $\hat{T}_{\rm ad}(1)=6.8718>\hat{T}(1)=1.295$ and the
  Hilsch efficiency is $\xi_{\rm H}=1.8241$. The pressure $p(x)$ is a
  decreasing function of $x$.
\label{f6}
}
\end{figure*}

We expectedly have
\begin{eqnarray}
  \label{eq:23}
\widehat{Q}(x_0)=-x_0\hat{T}'(x_0)>0, \qquad 
\widehat{Q}(1)=-\hat{T}'(1)>0,   
\end{eqnarray}
i.e. the heat enters from the inner boundary $x=x_0$ and leaves at the
outer boundary $x=1$; see (\ref{eq:999}). We see numerically that
$\hat{Q}_1\lesssim\hat{Q}_2$; see Fig.~\ref{f6}.

Now $\Delta\widehat W_\phi<0$ (rotating cylinders invest work), as we
discussed after (\ref{work_tangential}). But the radial external
forces do extract work, $\Delta\widehat{W}_r>0$ as much that the total
work is extracted [see Figs.~\ref{f6}, \ref{f7}]:
\begin{eqnarray}
  \label{eq:20}
\Delta\widehat{W}=\Delta\widehat{W}_r+\Delta\widehat{W}_\phi>0.
\end{eqnarray}
Moreover, the overall kinetic energy (see (\ref{energy})) also
increases, $\Delta\widehat{E}_{\rm kin}>0$ (albeit slightly, as seen
in Fig.~\ref{f6}) due to contribution
$\frac{\kappa}{2}(1-x_0^{2+2\kappa})$ of the vortex.

The most interesting aspect of this cooling scenario is that the
cooling efficiency (\ref{bars}) is larger than $1$ [see
Fig.~\ref{f6}]:
\begin{eqnarray}
  \label{eq:22}
  \xi \geq 1,   
\end{eqnarray}
i.e. the adiabatic process provides less cooling, since now the heat
transfer from the system is essential; see after
(\ref{eq:45}). Together with (\ref{eq:22}), the Hilsch efficiency
(\ref{hilsch}) is also larger than 1: $\xi_{\rm H}>1$; see Fig.~\ref{f6}.

Eq.~(\ref{eq:22}) is thermodynamically consistent. Recalling
(\ref{eq:26_1}, \ref{eq:433}), we see that the upper bound
(\ref{eq:45}) amounts to $\frac{x_0\hat{T}'(x_0)}{\hat{T}(x_0)}
-\frac{\hat{T}'(1)}{\hat{T}(1)}$. This expression is positive; cf.
(\ref{eq:23}). Hence it is possible to have (\ref{eq:22}) provided
that the entropy production is sufficiently small, which appears to be
the case, as confirmed numerically. We stress again that $\xi>1$ is
possible due to heat conductivity; cf. the discussion after
(\ref{eq:45}).

Due to (\ref{eq:26_1}) and (\ref{eq:22}), the entropy entering to the
system is larger than the one that leaves it [cf. (\ref{stokes})]:
\begin{eqnarray}
  \label{eq:25}
s(r_2)-s(r_1)=c_p\ln[1/\xi].
\end{eqnarray}

Thus we get that (without any investment of overall external work) the
temperature, stagnation enthalpy and entropy decrease, while the kinetic
energy increases. The outgoing fluid is more ordered, since not only
its thermal energy decreases, but also the kinetic energy increases.

\subsection{Physical mechanisms of the effect and cooling without
  vortical motion}

\begin{figure*}
\vspace{0.25cm}
\includegraphics[width=18.4cm]{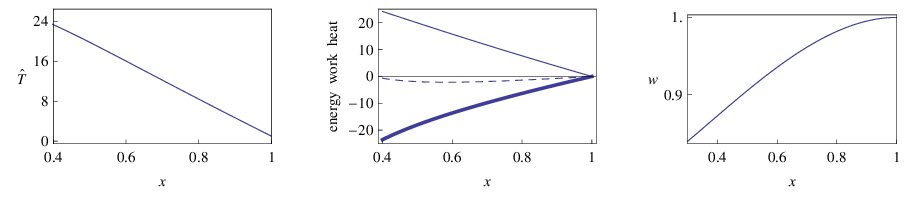}
\caption{Adiabatic cooling with outward ($c>0$) radial flow and without
  vortex motion ($v_\phi=0$). Dimensionless temperature $\hat{T}(x)$, 
  energy quantities and dimensionless radial velocity $w$ 
  versus dimensionless distance $x$; see Table~\ref{tab1}. \\
  The curves are obtained from numerical solution of
  (\ref{s2}, \ref{s4}) for $\kappa=1$, $b=1$, $\hat{c}_p=3.5$,
  $\beta=40$ ($\hat{T}'(1)=-36.5$), $\chi=10$ and $\hat{T}(1)=1$,
  $w(1)=1$, $w'(1)=0$. \\ 
  Left figure: $\hat{T}(x)$ versus $x=r/r_2$
  for $x\in [x_0,1]$, where $x_0=0.4$ (the dimensionless
  stagnation enthalpy $\hat{U}(x)$ behaves similarly). \\
  Middle figure: dimensionless energy
  $\widehat{E}(x)-\widehat{E}(1)$ (solid line), radial work
  $\widehat{W}_r(x)-\widehat{W}_r(1)$ (dashed line) and heat
  $\widehat{Q}(x)-\widehat{Q}(1)$ (bold line) versus $x$. \\
  Right figure: $w(x)$ versus $x$. Condition (\ref{peppy}) holds;
  cf. the data for Fig.~\ref{f6}. It also holds that $[w(x)/x]'<0$.\\
  The minimal dimensionless temperature
  is $\hat{T}(1)=1$ and $\hat{T}''(1)=6.4293$.\\
  Energetics: $\Delta \widehat{E}_{\rm kin}=-1.8809$,
  $\Delta\widehat{W}=\widehat{W}_r=0.6676$,
  $\widehat{Q}(x_0)=-x_0\hat{T}'(x_0)=12.9593$,
  $\widehat{Q}(1)=-\hat{T}'(1)=36.5$. \\
  Efficiency: $\hat{T}_{\rm ad}(1)=9.12373>\hat{T}(1)=1$ and the
  Hilsch efficiency is $\xi_{\rm H}=1.5719$. The pressure $p(x)$ 
  is a decreasing function of $x$.
\label{f7}
}
\end{figure*}

We saw around (\ref{eq:38}, \ref{kon}) that in order to get inward
flow cooling it is necessary to have an angular motion with the
viscous forces doing work of the proper sign. Here the physical
meaning of cooling can be clarified along the same lines. We get from
(\ref{s2}) and (\ref{energy}--\ref{heat})
\begin{eqnarray}
  0= \left[ \widehat{E}(x)+\widehat{Q}(x)+\widehat{W}_r(x)+
\widehat{W}_\phi(x)\right]'.
  \label{eq:388}
\end{eqnarray}
Now cooling implies that outgoing fluid has lower energy:
$\widehat{E}'(x)<0$. Possible necessary conditions for cooling is
provided by the heat conductivity: $\widehat{Q}'(x)>0$, and/or by the
work done via radial viscosity: $\widehat{W}_r'(x)>0$. Figs.~\ref{f6}
and \ref{f7} show that both these conditions hold. The contribution of
$\widehat{Q}'(x)>0$ is larger than that of $\widehat{W}_r'(x)>0$.

But the vortex contribution has the same sign as energy:
$\widehat{W}_\phi'(x)<0$; cf. with (\ref{kon}). Hence a similar
cooling scenario is also possible without vortex, i.e. for
$\hat{v}_\phi=0$ in (\ref{s2}, \ref{s4}). In fact, eliminating the
angular motion almost does not change the temperature profile in
Fig.~\ref{f6}.  The main difference with the above situation is that
the kinetic energy decreases, $\Delta E_{\rm kin}<0$, since the vortex
motion is now absent.

Note that to eliminate the vortex from equations of motion, one should
take $v_\phi=0$ in (\ref{s2}, \ref{s4}), and to suppress the last
factor $ -\alpha(2+\kappa)+2+\frac{\kappa}{2}$ in (\ref{eq:15}), as
well as $-(\alpha(2+\kappa)-2)^2$ in (\ref{on}). Definitions
(\ref{eq:34}, \ref{eq:34.0}, \ref{eq:34.1}) of dimensionless
parameters still apply, where now $v_2=v_{2,0}$ is an arbitrary
characteristic velocity; see section \ref{depo}. It drops out from
(\ref{s2}, \ref{s4}).

\comment{ This is most probably wrong:

Fig.~\ref{f6} shows that the cooling effect increases upon
increasing $w(1)$. (The constraint (\ref{eq:31}) is respected; see
(\ref{ona}).) The reason of this increasing is that the contribution
from the radial viscous work gets stronger; cf. (\ref{eq:388}). }

A simple analytical description of the temperature profile can be
obtained from (\ref{s2}) assuming that the change of $w(x)$ can be
neglected. (But note that the change of $w(x)$ within (\ref{s4})
cannot be neglected.) Taking in (\ref{s2}) $\hat{v}_\phi=0$,
$w(x)=w(1)$ and $w'(x)\to 0$, we get
\begin{gather}
\label{ho}
\hat{T}(x)-\hat{T}(1)=
\gamma [1-\frac{1}{x^2}]
+[2\gamma-\hat{T}'(1)]\,
\frac{1-x^b}{b}, \\
\gamma\equiv\frac{w(1)^2(4+\kappa)}{2(2+b)}.
\end{gather}
Now $\hat{T}'(x)=0$ is solved as $x_{\rm
  max}^{b+2}=\frac{2\gamma}{2\gamma -\hat{T}'(1)}$. This solution
exists only for $\hat{T}'(1)<0$ (recall that $b>0$) and it is a
maximum of $\hat{T}(x)$. Hence one should take $x_0>x_{\rm max}$ in
order to get a monotonic decrease of temperature $\hat{T}(x)$ from
$x=x_0$ till $x=1$.


One feature of this cooling scenario is that once $w(1)$ (the
boundary condition for the outward flow) decreases, the solution of
hydrodynamic equations (\ref{s2}, \ref{s4})) ceases to exist below a
certain critical value of $w(1)$, because now $w(x_0)$ becomes
negative. Recall from (\ref{eq:4}) that a negative $w(x_0)$ for $c>0$
is not acceptable, since it would mean a negative mass density $\rho$.
Thus, if $w(1)$ is decreased, then $c$ also has to
decrease, to keep the solution physical.


Generally, the magnitude (\ref{eq:188}) of the cooling increases upon
decreasing the radial flow $c$, i.e. for $c\to 0$. This can be seen
from (\ref{ho}) or from Figs.~\ref{f7}. But this limit $c\to 0$ is not
useful, since it diminishes the cooling power $b(
\hat{T}(x_0)-\hat{T}(1))$.

\subsection{Geometric aspects of the cooling effect}

Note that the present cooling scenario (without vortex) is specific
for the cylindrical geometry. Its traces are seen in 1d (plane
geometry), but the cooling as such is negligibly weak there; see
Appendix \ref{1d}.

To understand the origin of this effect, let us take the situation,
where all the velocities vanish (hence $c=0$).  Due to the cylindrical
geometry, (\ref{bora}) shows that there is still a non-trivial
stationary temperatures profile, $\ec= -\lambda rT_0'$ that reads in
terms of the dimensionless temperature $\hat{T}$ and dimensionless
length $x=r/r_2$ ($x_0\leq x\leq 1$):
\begin{eqnarray}
  \label{eq:57}
  \hat{T}_0(x) -   \hat{T}_0(1)=  \hat{T}'_0(1)\ln x.
\end{eqnarray}
Now it should be clear that (\ref{eq:57}) does not describe as such
any cooling effect. Indeed, for $\hat{T}'_0(1)\leq 0$
we should put a thermally isolated wall at $x=1$ (to avoid assuming
the existence of even colder temperatures), which leads us to 
$\hat{T}'_0(1)= 0$, and hence to a constant temperature profile. 

However, there is a relation between 
(\ref{eq:57}) and the temperature profiles obtained above for $c>0$ and
illustrated in Figs.~\ref{f6}, and \ref{f7}:
\begin{gather}
  \label{eq:58}
  \hat{T}_0(x) >   \hat{T}(x), \\
  {\rm if}~~~
  \hat{T}'_0(1)=\hat{T}'(1)<0 ~~~ {\rm and}~~~
  \hat{T}_0(1)=\hat{T}(1),
  \label{eq:588}
\end{gather}
where we note from (\ref{eq:57}) that $\hat{T}''_0(1)=-\hat{T}'_0(1)$,
i.e. for $\hat{T}'_0(1)=\hat{T}'(1)<0$, $\hat{T}_0''(1)$ agrees with
the boundary condition (\ref{eq:99}) required for cooling.

Eqs.~(\ref{eq:58}, \ref{eq:588})|which we verified numerically|show
that the reported cooling effect is the modification of the formal
temperature profile (\ref{eq:57}) to the physical, situation with
$c>0$. In the 1d case (plane geometry) the general zero-velocity
temperature profile (the analogue of (\ref{eq:57})) is just $T_0={\rm
  const}$, which explains why the cooling effect is negligible there;
see Appendix \ref{1d}.

\comment{Thoughts:

-- try to put initial conditions at $x_0$. If $\hat{T}'(1)<0$ and
$\hat{T}'(1)>0$ select two regimes, what a regime will be selected by 
$\hat{T}(x_0)'<0$?

-- there is a difference between Couette temperature distribution and
what is employed for the adiabatic cooling of the out-going
fluid. The difference is in the behavior of the pressure; for the
Couette flow it always decreases towards the center. A related point
is the ``hydro-static'' equilibrium does not work in the out-going
flow. What substitutes this equilibrium? Why specifically it does not
work? 

}

\section{Summary}

We worked out a tractable model for describing gasodynamic
cooling. The model extends the standard Couette flow between two
coaxial cylinders by adding there a radial flow (hence demanding that
cylinders are permeable). Only the radial dependence of relevant
quantities is retained and the axial flow is neglected. 

The model accounts for viscosity, heat-conductivity and
compressibility; see section \ref{notations}. They are {\it generally}
important for gasodynamic cooling, and there are at least two reasons
for keeping each of them in the description. Viscosity is to be
retained, (first) since we should achieve cooling also in terms of the
stagnation enthalpy (as observed experimentally \cite{savino}), and
(second) since due to turbulence the actual viscosity is much larger
than its molecular value. Heat-conductivity is to be considered
simultaneously with viscosity, since the Prandtl number of air is
close to one. It is also important for ensuring boundary conditions of
cooling. Compressibility is needed, because we need the proper
relation between fluid mechanics and thermodynamics, and also because
the involved angular velocities are sonic.

The emphasize of our study is not so much in describing details of the
vortex cooling effect as observed in experimental examples of vortex
tubes, but rather in showing how a hydrodynamic model can account for
cooling via specific boundary conditions (see section
\ref{definition}), and how already the simplest model can provide new
(and thermodynamically consistent) predictions for cooling with
efficiency larger than $1$.

We show that the model predicts a vortex cooling effect for an inward
radial flow; see section \ref{inward}. Though the general cause of
cooling is in the pressure gradient that drives the flow, the local
cause is related to the work done by viscous forces. The cooling
effect comes in two versions|adiabatic and isothermal|that are closely
related, but differ from each other by the boundary conditions. In
several ways the obtained cooling effect is similar to what was
experimentally seen in Ref.~\cite{savino} for a short uniflow vortex
tube. In accordance with experimental results, the model predicts that
the efficiency of vortex cooling is generically smaller than 1 [see
section \ref{jancooling}], though the concrete values for the
efficiency and for the magnitude of cooling are lower than what was
observed for best vortex tubes.

The model predicts as well a cooling effect that was (to our
knowledge) so far not observed experimentally; see section
\ref{semi}. This effect is realized for an outward flow and it does
not need an angular (vortical) motion.  Its cooling efficiency is larger than
one, i.e. for the given gradient of pressure, this cooling is more
efficient than the adiabatic (i.e. entropy conserving) thermodynamic
process. This cooling effect is consistent with the second law, and it
is possible due to heat-conductivity. It has partly a geometric
origin, since it is negligible for the plane geometry.

There is an experimental report on the Hilsch efficiency being larger
than 1 for a counter-flow tube (where only a part of the air is
cooled) \cite{martyn}. However, this result was was not reproducible
in \cite{guliguli}. According to the author of Ref.~\cite{guliguli},
the report concerned externally cooled vortex tubes; cf. our
discussion in section \ref{definition}. Hence we conclude by stressing
that cooling with an efficiency larger than one is an open problem.


\appendix

\section{Ideal gas thermodynamics}
\label{idol}
We briefly recall ideal-gas formulas as applied in hydrodynamics.
Thermodynamic relations of hydrodynamics are written for extensive
quantities are divided by the overall number $N$ of involved particles
and by the mass $m$ of a single particle. Thus the extensive ideal
gas entropy
\begin{eqnarray}
  S=k_{\rm B} C_v\ln [p]+k_{\rm B} C_p\ln [V/(Nm)],
 \end{eqnarray}
 where $C_v$ and $C_p$ are heat-capacities and $V$ is the volume,
 becomes
\begin{eqnarray}
s=\frac{  S}{Nm}=\frac{k_{\rm B}}{m}\frac{ C_v}{N}
\ln [p]+\frac{k_{\rm B}}{m}\frac{ C_p}{N}\ln [V/(Nm)]\nonumber\\
=\frac{k_{\rm B}N_A}{N_Am}\hat{c}_v \ln [p]
-\frac{k_{\rm B}N_A}{N_Am}\hat{c}_p \ln [\rho],
\label{ad1}
 \end{eqnarray}
 where $\rho=Nm/V$ is the mass density, $N_A$ is the Avogadro number,
 and where $\hat{c}_v$ and $\hat{c}_p$ are dimensionless numbers of
 order one:
\begin{eqnarray}
\hat{c}_p-\hat{c}_v=1.  
\end{eqnarray}
After denoting 
\begin{eqnarray}
  \label{eq:36}
k_{\rm B}N_A=R, \qquad N_Am=\mu,
\end{eqnarray}
where $R=8.314$ J/K is the gas constant, and $\mu$ is the molar mass,
(\ref{ad1}) reads:
\begin{eqnarray}
\label{grno}
&&  s={c}_v \ln [p] -{c}_p \ln [\rho], \\
&&  c_v=(R/\mu)\,\hat{c}_v, ~~   c_p=(R/\mu)\,\hat{c}_p.
\end{eqnarray}
The full entropy $S$ (and similarly other extensive quantities) is
obtained as $S=\int_V \d^3 x\, \rho\, s$.  Noting that the temperature
$T$ is measured in Kelvins, the ideal gas equation of state $pV=k_{\rm
  B}NT$ becomes
\begin{eqnarray}
\label{mrdo}
  p=(R/\mu)\,\rho T.
\end{eqnarray}
For purposes of dimensionless analysis, we write 
(\ref{grno}) as
\begin{eqnarray}
  \label{pox}
s=c_p\left(
-  \frac{1}{\hat c_p}\ln [p]+\ln [T] -\ln\left[
    \frac{\mu}{R}  \right] \right).
\end{eqnarray}
Eq.~(\ref{pox}) implies that if the pressure and temperature
adiabatically (i.e. for a constant entropy) change as $p\to p'$ and
$T\to T'$, then
\begin{eqnarray}
  {T'}/{T}=\left({p'}/{p}\right)^{1/\hat{c}_p}.
\label{bubu}
\end{eqnarray}

\section{Vortex flow without radial motion (Couette flow) }
\label{couette}

Consider the distribution of temperature inside of the vortex
(\ref{eq:6}) when the radial motion is absent. This is one of standard
problems of hydrodynamics (the Couette flow) and it is studied in many
places; see e.g. \cite{landau,dorfman}. We reconsider this problem
here, because we want to understand why specifically this situation
does not contain any interesting stationary cooling scenario (contrary
to remarks given in \cite{dorfman}). For
\begin{eqnarray}
  \label{gor}
  v_r=c=w=\kappa=b=0,
\end{eqnarray}
we get $(\frac{\kappa}{2}+1) \hat{v}_\phi^2 - x
\hat{v}_\phi\hat{v}'_\phi +b\hat{T}-x\hat{T}' -\beta=0$ from
(\ref{s2}). This equation integrates and determines temperature inside
of the vortex ($x\equiv r/r_2$)
\begin{eqnarray}
&&\hat{T}(x)=\hat{T}(1)-(1-\alpha)^2(\frac{1}{x^2}-1)\nonumber\\
&&+(\hat{T}'(1)-2(1-\alpha)^2)\ln x, ~~~~
x_0\equiv\frac{r_1}{r_2}\leq x\leq 1,~~
\label{hu1}  
\end{eqnarray}
where we employed (\ref{eq:15}, \ref{gor}) for expressing $\beta$ via
$\hat{T}'(1)$, and where $\alpha$ is given by (\ref{eq:6.1}) under
$\kappa=0$:
\begin{eqnarray}
  \label{eq:39}
\alpha = \frac{1-(v_1r_1)/(v_2r_2)}{1-(r_1/r_2)^{2}}.
\end{eqnarray}
Note that taking the inner radius $r_1$ to zero, $r_1\to 0$, does not
lead to anything interesting: in this limit we get $\alpha\to 1$,
(\ref{hu1}) implies $\hat{T}(x)= \hat{T}(1) + \hat{T}'(1)\ln x$, but
we have to assume also $\hat{T}'(1)=0$ for preventing the singularity
at $x\to 0$. Then $\hat{T}(x)$ does not depend on $x$. Hence $r_1$
should be kept finite.

Interesting (stationary) cooling scenarios are those, where the low
temperatures created inside of the fluid are not due to even lower
temperatures imposed on its boundary. In particular, if one of the
boundaries is left without thermal isolation|so there is an active
thermal bath working at this boundary|then the inside temperature
should be lower than the temperature of this bath.

Let us start with the case, where no thermal isolation is imposed for
both boundaries. Then $\hat{T}(x)$ should have a local minimum at some
$x\in (x_0,1)$. Eq.~(\ref{hu1}) shows that there is only one solution
$x=x_{\rm max}$ of $\hat{T}'(x)=0$:
\begin{eqnarray}
  \label{eq:13}
  x_{\rm max}=\sqrt{\frac{2(1-\alpha)^2}{2(1-\alpha)^2-\hat{T}'(1)}}.
\end{eqnarray}
If $0<x_{\rm max}<1$ (for which it is necessary and sufficient that
$\hat{T}'(1)<0$), then $x_{\rm max}$ is the local maximum (not
minimum) of $\hat{T}(x)$. Hence we get no cooling for this case.

Next, let us thermally isolate the outer boundary: $\hat{T}'(1)=0$. We
get from (\ref{hu1}):
\begin{eqnarray}
  \hat{T}(x)-
  \hat{T}(1)=-(1-\alpha)^2\left(\frac{1}{x^2}-1-\ln[\frac{1}{x^2}]
\right).
\end{eqnarray}
Then $\hat{T}(x)$ is a monotonically increasing function of $x$,
i.e. the inside temperature $\hat{T}(x)$ is larger than the (inner)
bath temperature $\hat{T}(x_0)$. For a thermally isolated inner
boundary, $\hat{T}'(x_0)=0$, $\hat{T}(x)$ is a monotonically
decreasing function of $x$ [see (\ref{eq:13})], i.e. again we get no
interesting scenarios of cooling. There are no solutions when both
boundaries are thermally isolated; see (\ref{hu1}). 

\comment{
The vortex pressure is found directly from (\ref{eq:1}):
\begin{eqnarray}
  \label{eq:17}
p(x) = p(1)
\exp\left[-
\frac{\mu\lambda}{R\eta}\int_x^1 \frac{\d y\,\hat{v}_\phi^2(y)}{y\hat{T}(y)}
\right].
\end{eqnarray}
Hence the pressure $p$ is a monotonously increasing function of $x$.
}

The absence of interesting cooling scenarios is confirmed by looking
at the total work produced by external forces that rotate the
cylinders. It reads from (\ref{vagr}) [with $\kappa=0$]:
\begin{eqnarray}
\widehat{W}_\phi
=2(1-\alpha)^2(1-\frac{1}{x_0^2}) <0.
\label{agr}
\end{eqnarray}
The negativity of (\ref{agr}) means that the work is invested
externally and dissipated for overcoming the viscous forces. This work
leaves the system as heat.

Thus three regimes are impossible for the considered Couette flow:

-- It cannot cool the fluid isothermally, i.e. when both boundaries
are kept at the same temperature.

-- It cannot cool the fluid adiabatically: no regime exists when one
boundary is thermally isolated, while another one is subject to a
thermal bath, and it is demanded to get the fluid colder than the
active bath temperature. Hence low-temperatures present in the system
according to (\ref{hu1}) do not constitute any non-trivial cooling:
they are due to the low-temperature bath present at one of
boundaries. Put differently, for the Couette flow the active bath is
the one with the lower temperature.

The above two conclusions seem to hold rather generally for stationary
hydrodynamic systems without mass flow, though we so far did not get a
general argument for their validity. At any rate they hold for the
(generalized) Couette (sometimes also called Taylor-Dean) flow, where
the fluid is subject to azimuthal driving with a volume force
$\vec{f}$, where only the $\phi$-component $f_\phi(r)$ of $\vec{f}$ is
non-zero, but it is an arbitrary function of $r$.

-- The Couette flow cannot also function as a heat-engine,
since|irrespectively of the values of $T(r_2)$ and $T(r_1)$|the work
is always dissipated; see (\ref{agr}).

\section{1d example of weak adiabatic cooling}
\label{1d}

Consider a 1d flow (from left to right) between two permeable plates
separated by distance $L$. Continuity of mass leads to
\begin{eqnarray}
  \label{eq:33}
  \rho v =c_1={\rm const}.
\end{eqnarray}

1d Navier-Stokes and energy equations read in dimensionless form 
\begin{gather}
  \label{eq:29}
\kappa \hat{v}^2(x)+\frac{b}{\hat{c}_p}\hat{T}(x)
  -(\chi+\frac{4}{3})\hat{v}(x)\hat{v}'(x)=\gamma \hat{v}(x),\\
  \label{eq:299}
\frac{\kappa}{2} \hat{v}^2(x)+b \hat{T}(x)   
-(\chi+\frac{4}{3})\hat{v}(x)\hat{v}'(x)-\hat{T}'= \beta,
\end{gather}
where $0\leq x\leq 1 $, $L$ is the distance between two plates,
$\beta$ and $\gamma$ are constants, and we introduced the following
dimensionless parameters:
\begin{eqnarray}
  \label{32}
&& \hat{T}=\frac{\lambda }{v_2^2\eta}T, ~~\hat{v}(x)=\frac{v(x)}{v(L)} \\
&& b=\frac{c_1c_pL}{\lambda}, ~~~~ \kappa=\frac{c_1L}{\eta}, \\
  \label{34.1}
&& \chi=\frac{\zeta}{\eta}, ~~~ \hat{c}_p=c_p(\mu/R). 
\end{eqnarray}
The constants $\beta$ and $\gamma$ can be expressed via (respectively)
$\hat{v}'(1)$ and $\hat{T}'(1)$:
\begin{eqnarray}
  \label{eq:35}
&&  \gamma=\kappa+ \frac{b}{\hat{c}_p}\hat{T}(1)
  -(\chi+\frac{4}{3})\hat{v}'(1), \\
&& \beta=\frac{1}{2}\kappa+ b\hat{T}(1)
  -(\chi+\frac{4}{3})\hat{v}'(1)-\hat{T}'(1).
\end{eqnarray}
We obtain from (\ref{32}--\ref{eq:35}):
\begin{gather}
  \label{osho}
  \hat{T}''(1)=b\hat{T}'(1)(1-\frac{1}{\hat{c}_p})
-(\chi+\frac{4}{3})\hat{v}'(1)^2
+ \frac{b}{\hat{c}_p} \hat{T}(1)\hat{v}'(1).
\end{gather}

Let us now look at conditions for adiabatic cooling. Now $c_1>0$
(hence $b>0$ and $\kappa>0$) and $\hat{v}>0$ from (\ref{eq:33}). Hence
we look for
\begin{eqnarray}
  \label{eq:37}
\hat{T}(0)>\hat{T}(1), \qquad \hat{T}''(1)>0.  
\end{eqnarray}
The temperature profiles appear to be monotonic so that the first
condition in (\ref{eq:37}) can be written as $\hat{T}'(1)<0$.  Then
the first and second term in the right-hand-side of (\ref{osho}) are
negative. Hence $\hat{T}''(1)>0$ can be satisfied only due to
sufficiently large $\frac{b}{\hat{c}_p} \hat{T}(1)\hat{v}'(1)>0$. This
implies limitations on $T(1)$ (which cannot be sufficiently small) and
on $|\hat{T}'(1)|$ (which cannot be sufficiently large). Eventually,
the adiabatic cooling appears to be a relatively small effect,
though it is still possible in this model.  For example, under
$\chi=10$, $b=\kappa=10$, $\hat{c}_p=3.5$, $\hat{T}(1)=1$,
$\hat{T}'(1)=-0.1$ and $\hat{v}'(1)=0.1$ (we have $\hat{v}(1)=1$ by
definition) we get for the cooling magnitude
$[\hat{T}(0)-\hat{T}(1)]/\hat{T}(0)=0.037$.

\section{Potential vortex}
\label{potential_vortex}

Another familiar type of vortex in (\ref{eq:6}, \ref{eq:6.0}) is:
\begin{eqnarray}
  \label{fort}
\alpha=0 ~~{\rm or}~~  \frac{v_1}{v_2} = \frac{r_2}{r_1}.
\end{eqnarray}
Eqs.~(\ref{lopes}, \ref{gomes}) imply
\begin{eqnarray}
  \hat{T}(x)-\hat{T}(1)&=&[\hat{T}'(1)- \frac{\kappa+4}{b+2}
 ]\,\,\frac{x^b-1}{b}\nonumber\\
&+&\frac{\kappa+4}{2(b+2)} (1-x^{-2}).
\end{eqnarray}
Now $\hat{T}'(x)=0$ is solved as
\begin{eqnarray}
  x^{-2-b}=1-\frac{(b+2)T'(1)}{(w^2(1)+1)(\kappa+4)}.
\end{eqnarray}
Hence the minimum of $\hat{T}(x)$ for $\hat{T}'(1)>0$ can exist only for 
\begin{eqnarray}
  \kappa<-4,  
\end{eqnarray}
i.e. only for radially inward flowing fluid ($c<0$). 

Now the isothermal cooling is driven by the work done for rotating
cylinders. Eqs.~(\ref{vagr}, \ref{fort}) imply
\begin{eqnarray}
  \label{eq:28}
  \widehat{W}_\phi=2(1-x_0^{-2})<0,
\end{eqnarray}
while the kinetic energy change is now larger than zero; see
(\ref{maga}, \ref{fort}). Due to this, $\hat{U}(x_0)>\hat{U}(1)$.

In other respects the two scenarios of cooling (quasi-solid and
potential) are similar to each other: both have roughly the same
magnitude, both need small radial velocities and both have efficiency
$\xi$ smaller than $1$.

\comment{
\section{Non-rotating (no vortex) situation}
\label{no-vortex}

For two non-rotating cylinders $v_\phi(r)=0$.  This case is described
by the following equations [cf. (\ref{s2}, \ref{s4})]:
\begin{gather}
\label{kokord1}
b\hat{T}-x\hat{T}' -\beta 
   +
  (\frac{\kappa}{2}+2)\frac{w^2}{x^2}  -(\chi+\frac{4}{3}) \frac{ww'}{x}
  =0,\\
 (\chi+\frac{4}{3})w''-(\kappa+\chi+\frac{4}{3})\frac{w'}{x} 
  +\frac{\kappa w}{x^2} -\frac{b\, x}{\hat{c}_p}\,(\hat{T}/{w})'
=0, 
\label{kokord2}
\end{gather}
where in (\ref{eq:34}--\ref{eq:34.1}) one can take $v_2=w(1)$; hence
the above equations are to be solved with the initial condition
$w(1)=1$. Eq.~(\ref{kokord2}) without the last term is solved exactly
$w(x)=\gamma x^{a_1} +(1-\gamma)x^{a_2}$, where $\gamma$, $a_1$ and
$a_2$ are constants, but this is typically a poor
approximation. Eq.~(\ref{kokord1}) can be solved exactly for
$w(x)=w(1)$, but this is a poor approximation again.
}


\comment{
\begin{figure*}
\vspace{0.25cm}
\includegraphics[width=12.4cm]{ran_1.eps}
\caption{Isothermal cooling with inward flow and quasi-solid vortex.  \\
  The same parameters as in Fig.~\ref{f1}, but with $w(1)=-0.005$,
  $w'(1)=0.01$,
  $\hat{T}_{\rm min}=\hat{T}(x_{\rm min})=1.13083$, $x_{\rm min}=0.50597$.\\
  The energy values: $\Delta \widehat{E}=-0.18874$,
  $\Delta\widehat{W}=0.35469$,
  $\widehat{Q}(x_0)=-x_0\hat{T}'(x_0)=0.07534$,
  $\widehat{Q}(1)=-\hat{T}'(1)=-0.09$. \\
  The pressure $\hat{p}(x)$ and $w(x)$ are monotonically increasing
  functions of $x\in [x_0,1]$. $\hat{T}_{\rm ad}(x_{\rm min})=0.9359$;
  see (\ref{eq:44}, \ref{eq:43}). The Hilsch efficiency $\xi_{\rm
    H}=0.13016$.  }
\label{f2}
\end{figure*}
}

\comment{
\begin{figure*}
\vspace{0.25cm}
\includegraphics[width=18.4cm]{ran_8.eps}
\caption{Adiabatic cooling with outward radial flow and without
  vortex motion ($v_\phi=0$). Dimensionless temperature $\hat{T}(x)$, 
  energy quantities and dimensionless radial velocity $w$ 
  versus dimensionless distance $x$; see Table~\ref{tab1}.  \\
  The curves are obtained from numerical solution of
  (\ref{s2}, \ref{s4})) under the same parameters as for
  Fig.~(\ref{f7}), but with the following changes.  Left figure:
  $w(x)$ versus $x=r/r_2$ for $w(1)=0.466$ and $\kappa=b=1$. It is
  seen that $w(x)$ gets small for $x\to x_0$: $w(x_0)=0.02$. \\
  Middle figure: the dimensionless temperature $\hat{T}(x)$ versus $x$
  for $w(1)=0.466$, but also $\kappa=b=0.1$. The cooling effect does
  improve as compared to Fig.~\ref{f7}, since $T(x_0)$ is larger,
  while $T(1)$ is the same. \\
  Right figure: $w(x)$ versus $x$ for $w(1)=0.466$, and
  $\kappa=b=0.1$. It is seen that $w(x_0)$ is not anymore small.
\label{f8}
}
\end{figure*}
}

\end{document}